 \documentclass[10pt,pra,aps,a4paper,oneside,twocolumn,superscriptaddress,showpacs]{revtex4-2}

\usepackage{enumitem}

\usepackage[english]{babel}
\usepackage[latin1]{inputenc}
\usepackage[T1]{fontenc}
\usepackage{lmodern}
\usepackage{ulem}
\usepackage{dcolumn}
\usepackage{subfigure}
\usepackage{footnote}
\usepackage{multirow}
\usepackage{amsmath,amsfonts,amssymb}

\usepackage{graphicx}  
\usepackage{latexsym}   
\usepackage{color}

\begin{document}

\title{Gain-controlled directional scattering in core-shell nanoparticles mediated by magnetic toroidal dipoles}

\author{\firstname{Tiago}  Jos\'e \surname{Arruda}}
\email{tiago.arruda@unifal-mg.edu.br}
\affiliation{Instituto de Ci\^encias Exatas (ICEx),
Universidade Federal de Alfenas (UNIFAL-MG), Alfenas 37133-840, Minas Gerais, Brazil}

\begin{abstract}
Toroidal dipole moments arise from poloidal current distributions and form a distinct class of electromagnetic excitations with unique near-field characteristics.
Using Lorenz-Mie theory, we show that interference between conventional magnetic and magnetic toroidal dipoles in core-shell nanoparticles produces Fano resonances and pronounced forward--backward scattering asymmetry.
By introducing optical gain in the dielectric core, we demonstrate that the toroidal mode can be selectively enhanced, enabling control of near-field confinement and far-field scattering directionality.
As the gain varies, we find that the system undergoes a continuous transition from suppressed backscattering to suppressed forward scattering through an intermediate regime of dominant magnetic-dipole radiation.
This dipolar scattering pattern is associated with a phase resonance of the magnetic toroidal dipole and a reversal of the poloidal current handedness.
These results identify gain-controlled toroidal excitations as a tunable mechanism for directional scattering in nanoscale systems.
\end{abstract}

\maketitle

The possibility of controlling the angular distribution of light scattered by particles with tailored optical properties has long been an important topic of research.
When the particle size is comparable to the wavelength, light scattering is commonly described in terms of electric and magnetic multipolar excitations, which account for a wide range of scattering phenomena~\cite{Arruda_PhysRevB109_2024,Grahn_NJPhys14_2012}.
By balancing different multipolar contributions in nanostructures, one can achieve directional scattering, including suppressed backward or forward radiation, often discussed in terms of generalized Kerker conditions~\cite{Luk_ACSPhot2_2015,Trigo_PhysRevLett125_2020}, as well as asymmetric spectral lineshapes associated with Fano resonances~\cite{Kivshar_RevModPhys82_2010,Luk_NatMat9_2010,Arruda_PhysRevA87_2013}.
These effects arise from interference between conventional electric and magnetic multipoles.

While conventional multipolar descriptions based on spherical electric and magnetic modes capture many scattering phenomena, they can obscure the role of complex current configurations governing the near-field response~\cite{Talebi_Nanoph7_2018}.
Among these, toroidal dipole excitations are central to interference-driven scattering phenomena, including anapole states~\cite{Kivshar_NatComm6_2015}, toroidal dipole-induced transparency and absorption~\cite{Miroshnichenko_LaserPhotRev9_2015,Jiang_OptExpress25_2017}, ideal magnetic-dipole scattering~\cite{Feng_PhysRevLett118_2017}, and unidirectional backscattering~\cite{Ge_OptExpress25_2017}.
Unlike conventional electric and magnetic multipoles, toroidal dipoles arise from poloidal currents forming closed loops along the meridians of a torus~\cite{Radescu_PhysRevE65_2002} and are associated with near-field energy confinement.
Although toroidal moments have long been recognized in nuclear and particle physics~\cite{Zeldovich_JExpTheorPhys33_1957}, their role in nanophotonics has gained attention only recently~\cite{Kaelberer_Science330_2010,Zheludev_NatMat15_2016}, and their contribution to far-field directionality remains only partially understood.

In spherical nanoparticles, toroidal multipole effects are typically inferred from local field distributions, since conventional multipole expansions implicitly encode toroidal contributions within spherical electric and magnetic coefficients~\cite{Muhlig_Metamat5_2011,Zhang_PhysRevA92_2015,Miroshnichenko_OptExp23_2015}. Recently, analytic expressions enabling explicit separation of electric and magnetic toroidal dipole contributions from the Lorenz-Mie coefficients were derived~\cite{Arruda_PhysRevB109_2024}, allowing direct investigation of toroidal electrodynamics in core-shell nanostructures. In gain-assisted nanoparticles, compensation of intrinsic losses enables selective amplification of specific modes and access to interference regimes unattainable in passive systems~\cite{Trigo_PhysRevLett125_2020,Shen_Nanophotonics6_2017}. Despite extensive studies of gain-assisted generalized Kerker effects and Fano resonances, the role of gain in controlling toroidal dipole excitations and their impact on scattering directionality remains largely unexplored.

Here, we investigate toroidal dipole excitations in gain-assisted core-shell nanoparticles as a route to actively control directional light scattering.
Using full-wave Lorenz-Mie theory with explicit separation of magnetic and magnetic toroidal dipoles~\cite{Arruda_PhysRevB109_2024}, we show that optical gain selectively enhances the toroidal response, producing Fano resonances and tunable forward-backward asymmetry.
Gain-induced toroidal-dipole phase evolution enables a transition from suppressed backscattering to suppressed forward scattering through an intermediate regime dominated by magnetic-dipole radiation. Unlike conventional gain-assisted Kerker schemes, where gain modifies the balance between electric and magnetic multipoles, the present mechanism is associated with a gain-controlled phase reversal of the magnetic toroidal contribution within the magnetic dipole channel.

\begin{figure}[htbp]
\includegraphics[width=0.9\columnwidth]{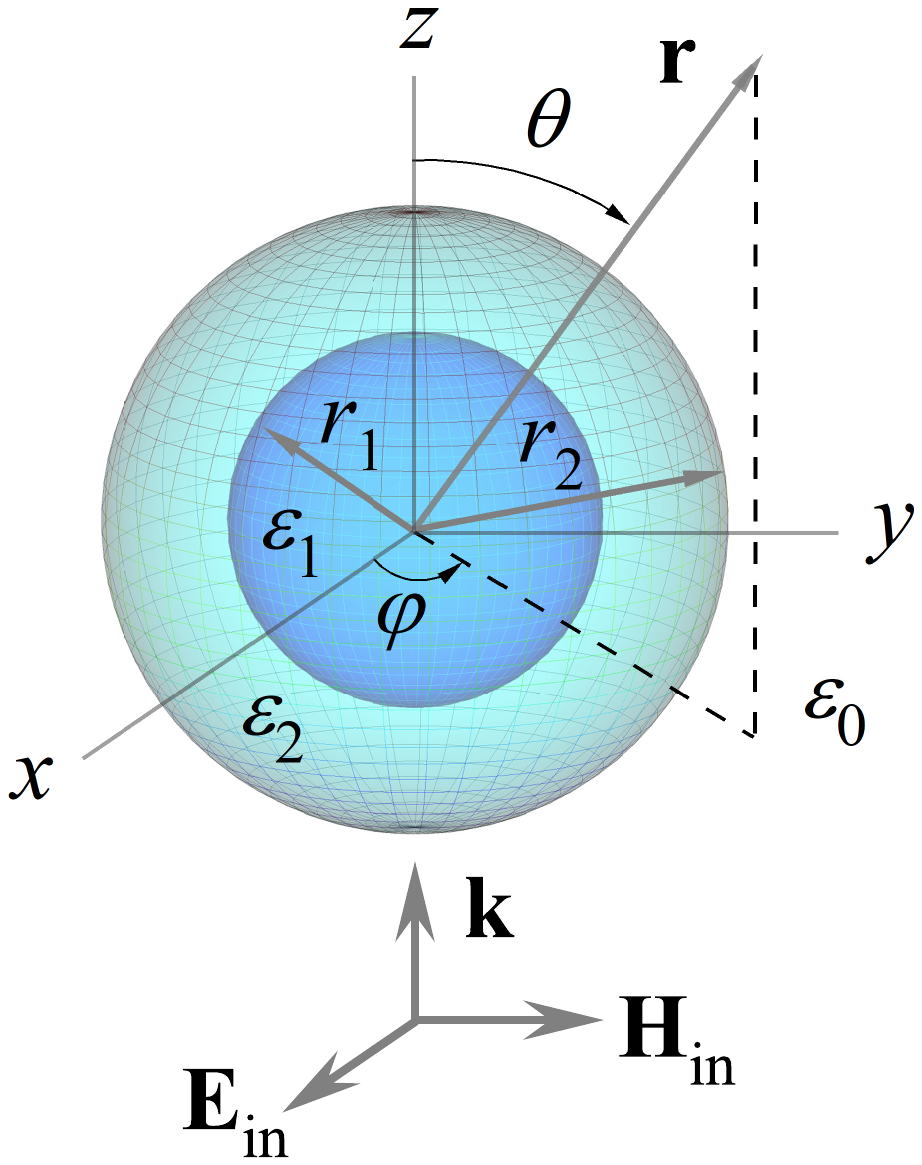}
\caption{Spherically symmetric core-shell nanosphere with inner radius $r_1$ and outer radius $r_2$, illuminated by a plane electromagnetic wave propagating along $\hat{\mathbf{z}}$. The permittivities of the core, shell, and surrounding medium are $\varepsilon_1$, $\varepsilon_2$, and $\varepsilon_0$, respectively.
 }\label{fig1}
\end{figure}

\begin{figure*}[htbp]
\includegraphics[width=.36\textwidth]{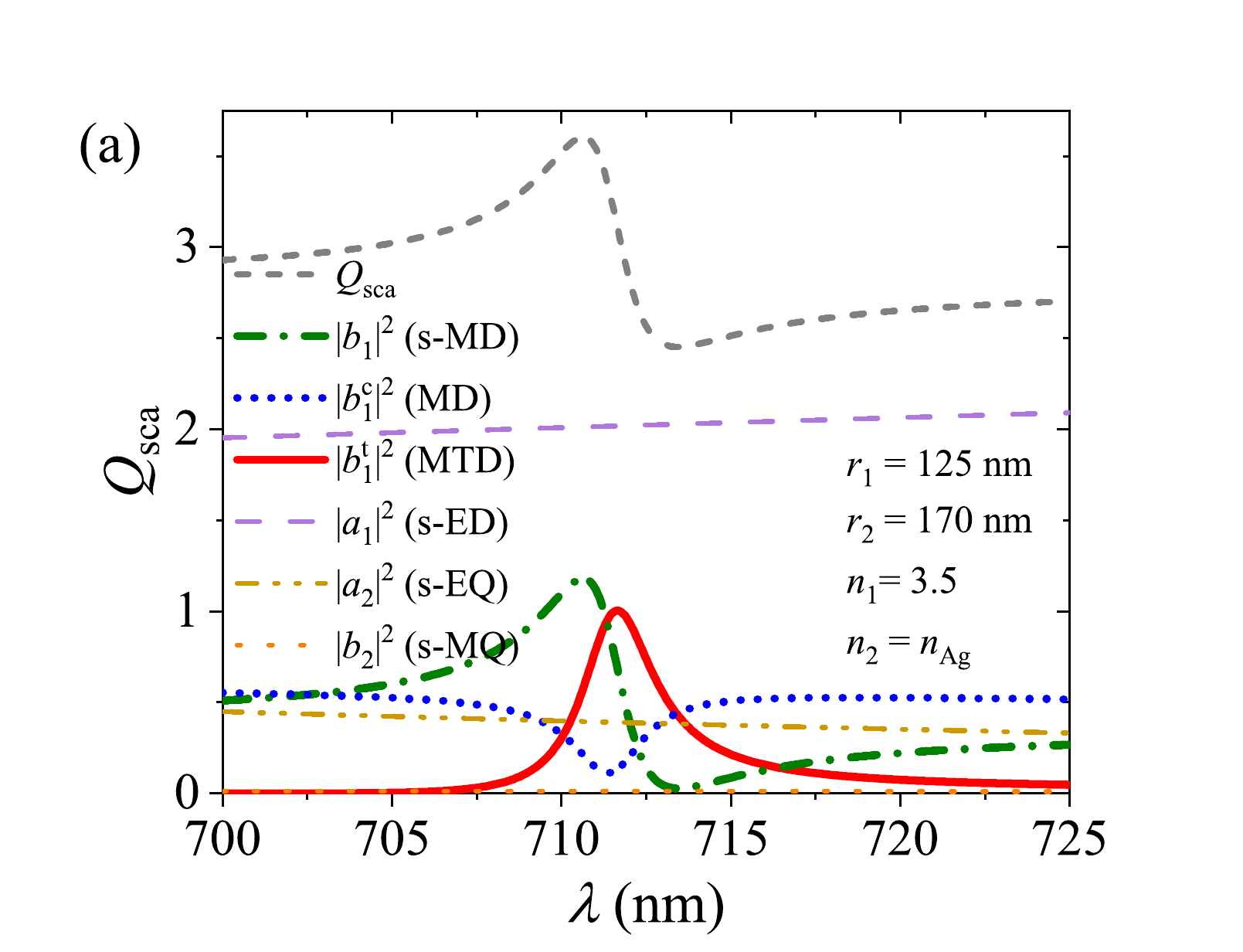}\hspace{-1cm}
\includegraphics[width=.36\textwidth]{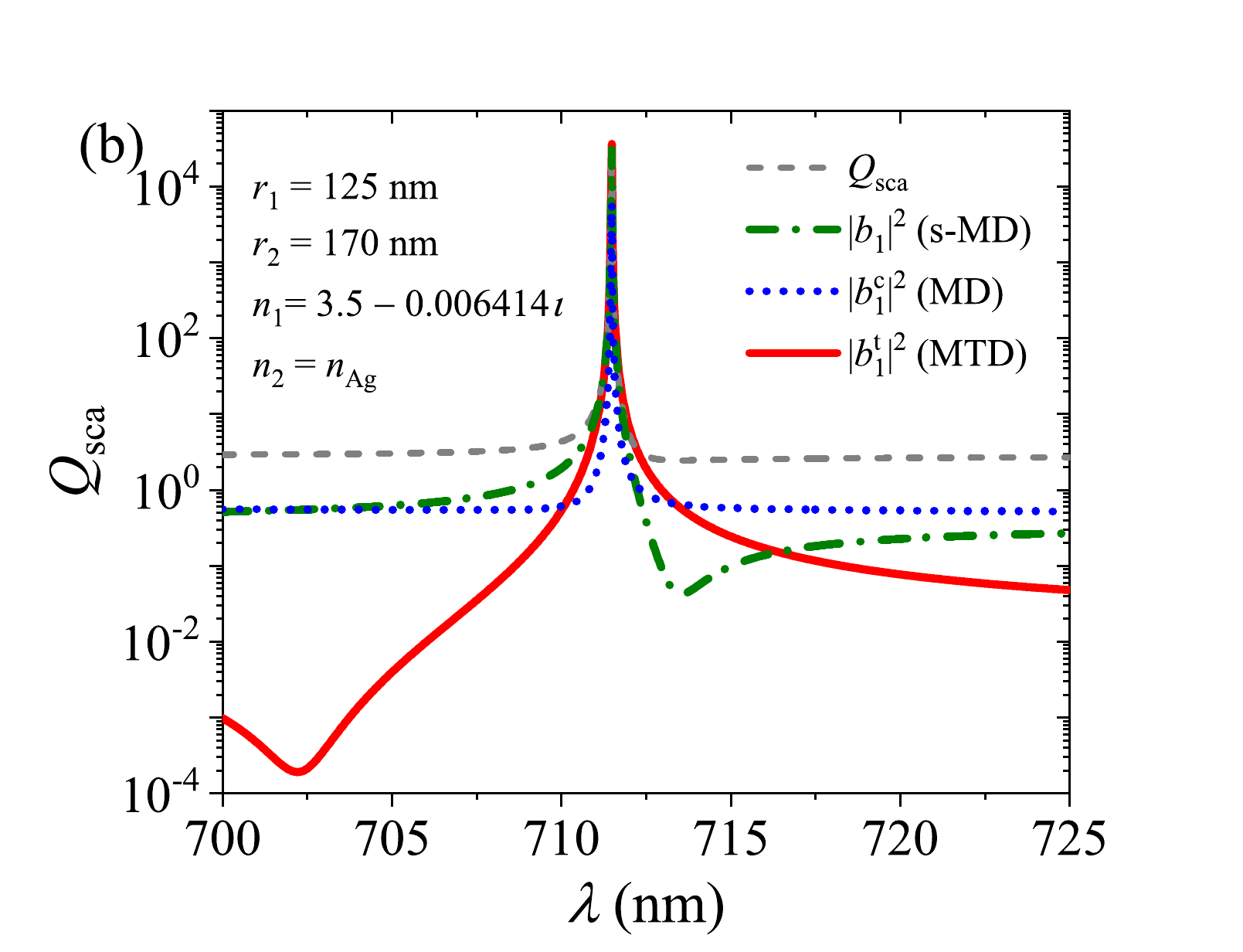}\hspace{-1cm}
\includegraphics[width=.36\textwidth]{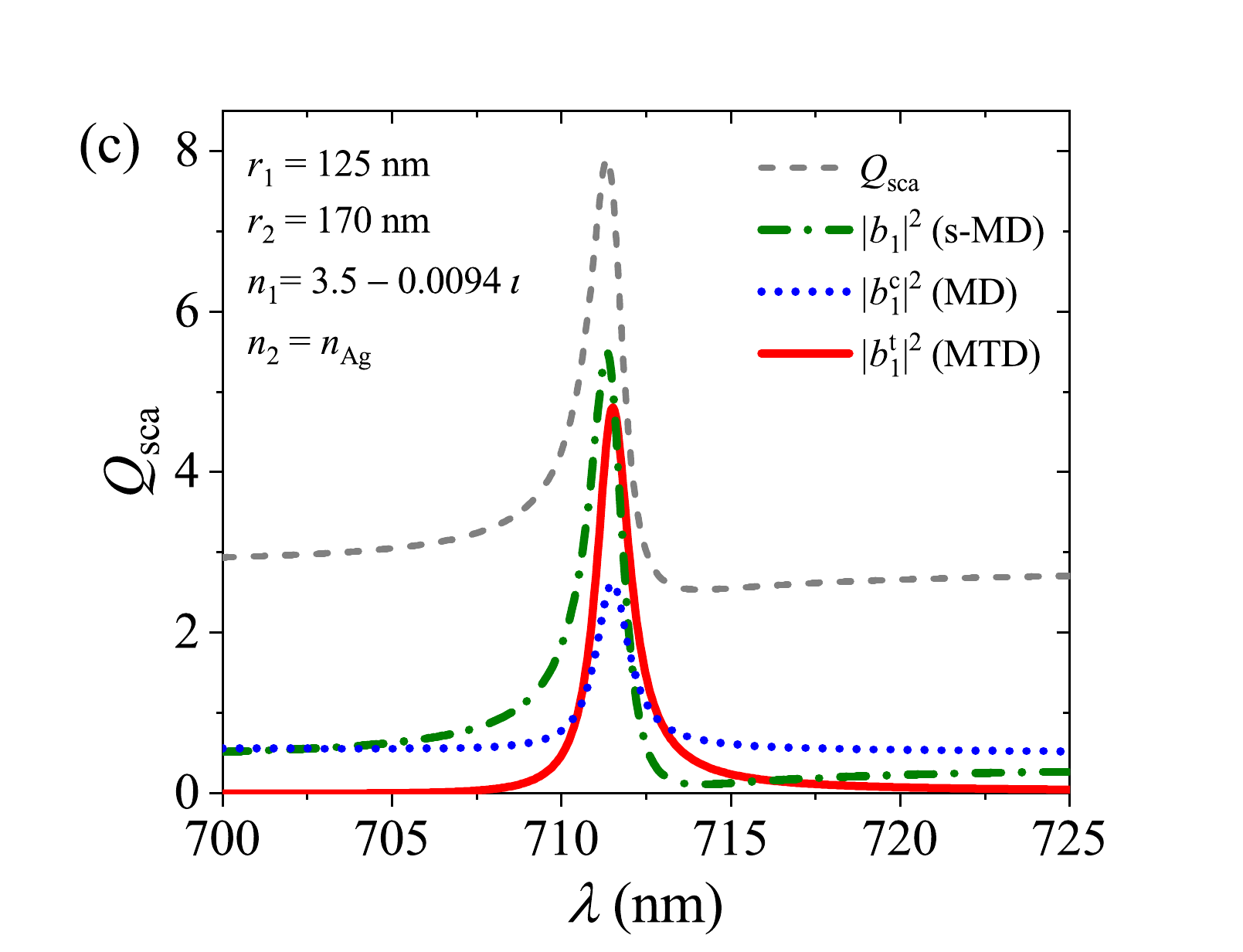}
\includegraphics[width=.35\textwidth]{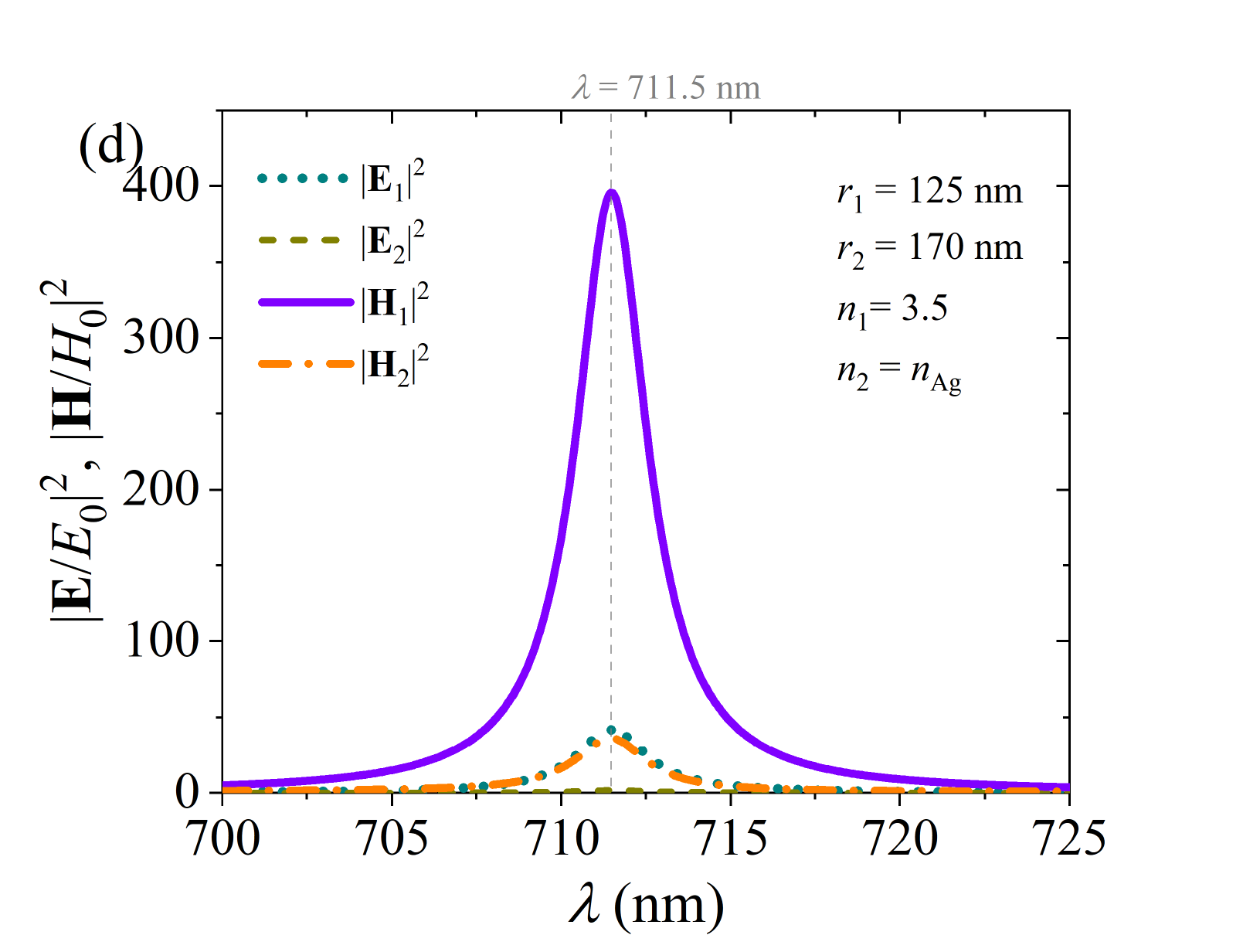}\hspace{-0.8cm}
\includegraphics[width=.35\textwidth]{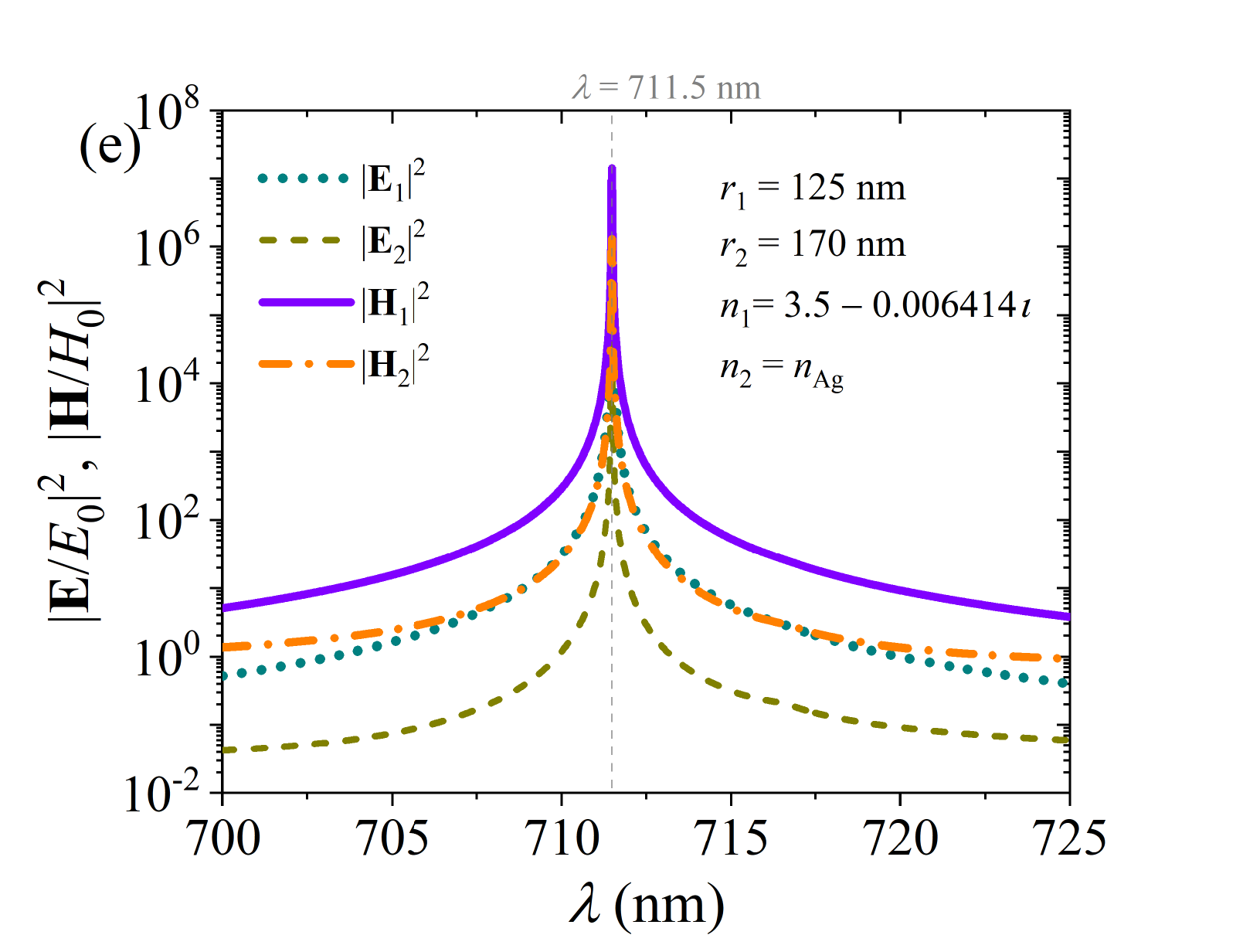}\hspace{-0.8cm}
\includegraphics[width=.35\textwidth]{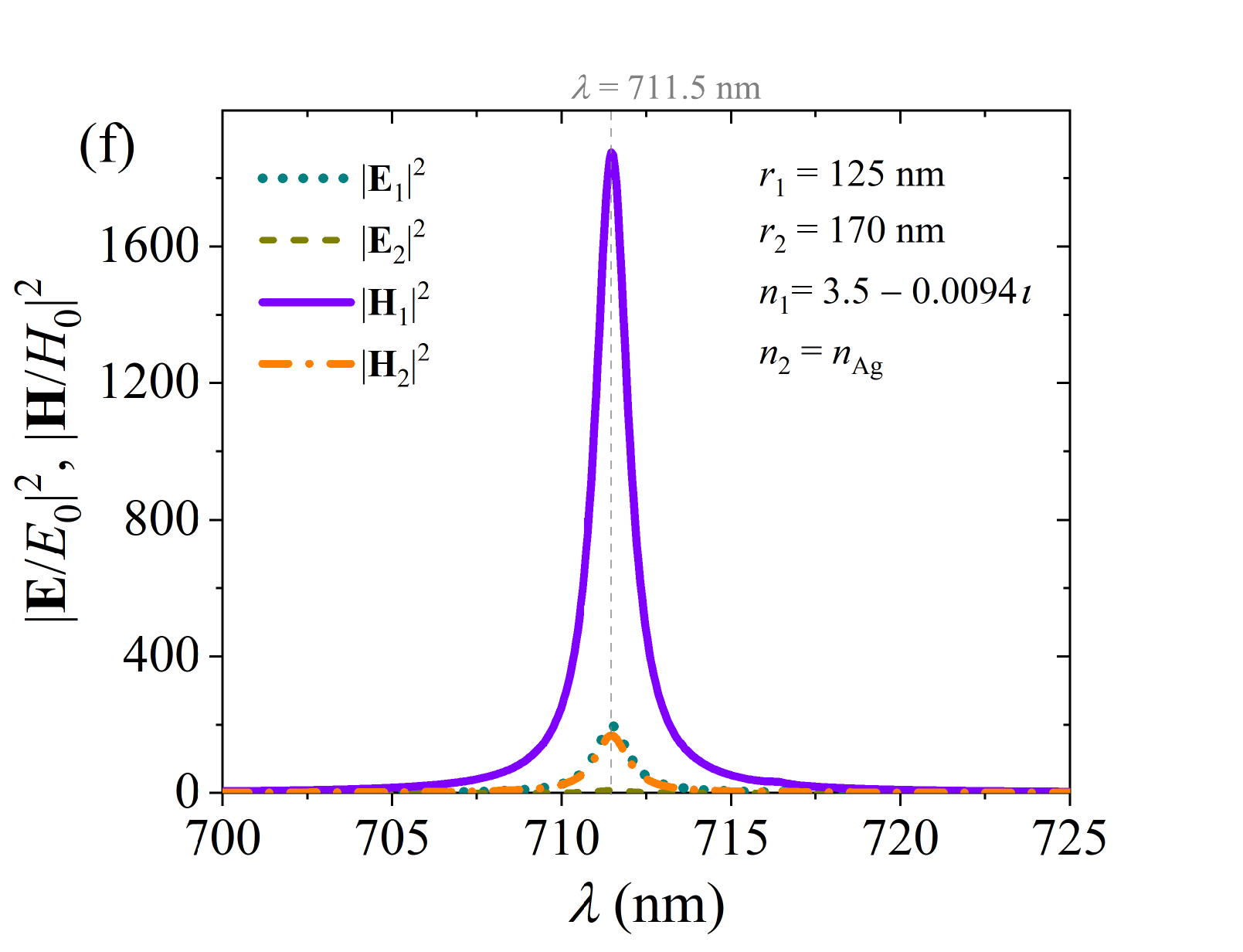}
\caption{Scattering by a gain-assisted dielectric nanosphere ($n_1=3.5-\imath\kappa$) of radius $r_1=125$~nm coated with an Ag nanoshell of outer radius $r_2=170$~nm as a function of wavelength. The scattering efficiency $Q_{\rm sca}$ is calculated using full-wave Lorenz-Mie theory. Individual multipolar contributions are identified by the Lorenz-Mie coefficients $a_1$ (s-ED), $b_1$ (s-MD), $b_1^{\rm c}$ (Cartesian MD), $b_1^{\rm t}$ (MTD), $a_2$ (s-EQ), and $b_2$ (s-MQ). Results are shown for (a) $\kappa=0$ (no gain), (b) $\kappa=0.006414$, and (c) $\kappa=0.0094$. Time-averaged electric and magnetic field intensities inside the core $(\mathbf{E}_1,\mathbf{H}_1)$ and shell $(\mathbf{E}_2,\mathbf{H}_2)$ are shown in (d)--(f) for the corresponding gain values. The vertical dashed line indicates the resonance wavelength $\lambda\approx711.5$~nm.
}\label{fig2}
\end{figure*}

 \begin{figure}[bpht]
\includegraphics[width=.45\textwidth]{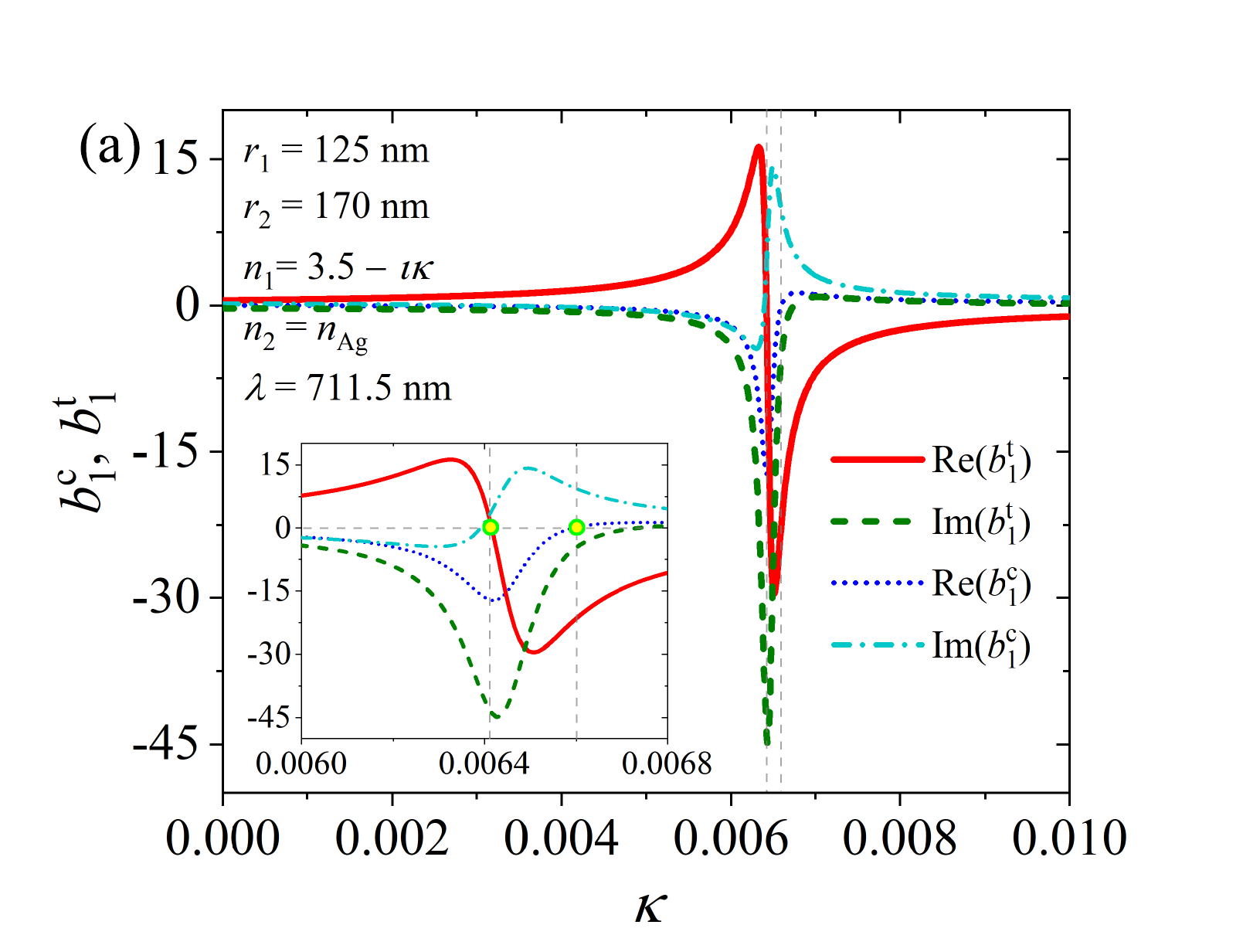}\vspace{-0.5cm}
\includegraphics[width=.45\textwidth]{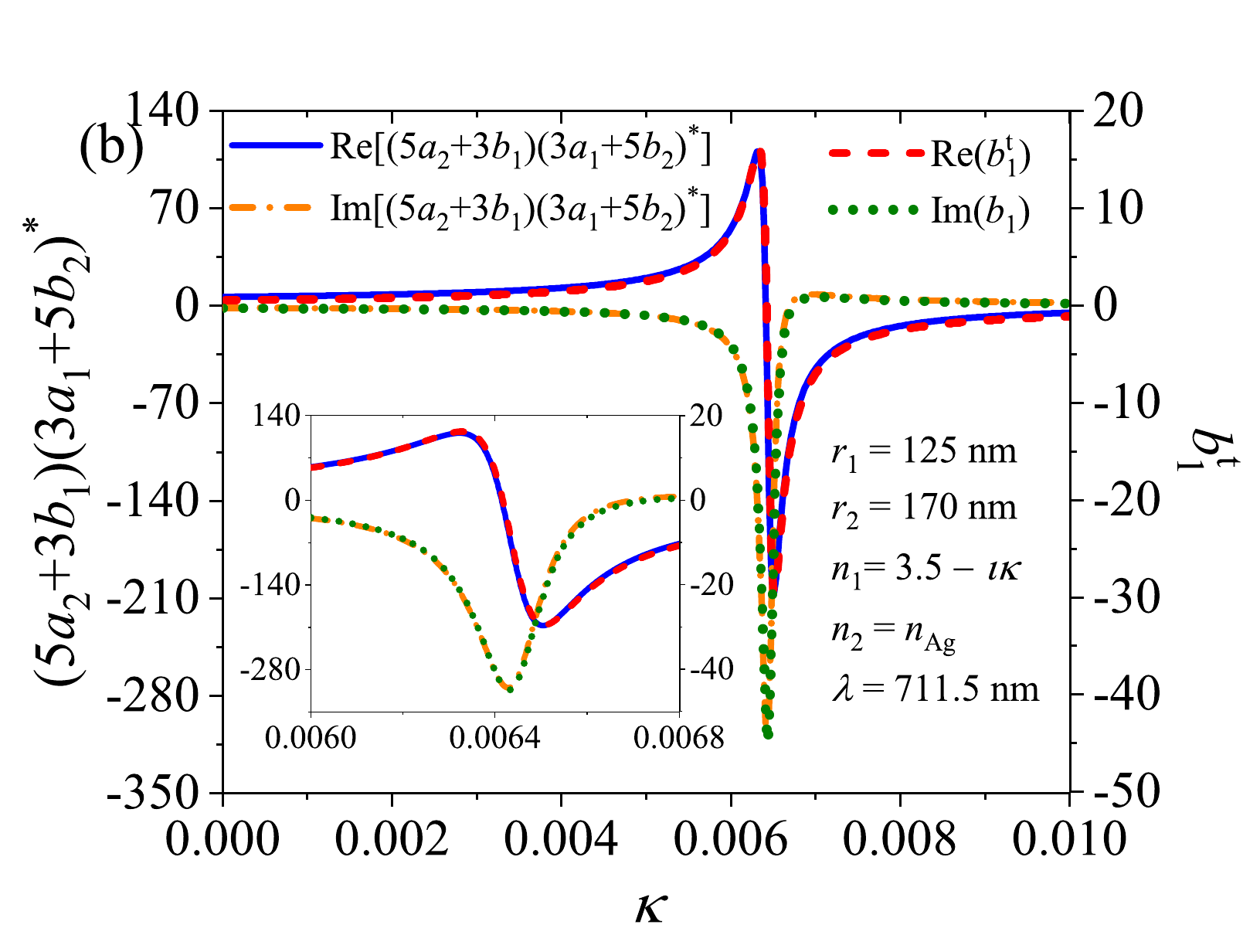}
\caption{(a) Real and imaginary parts of the Cartesian MD coefficient ($b_1^{\rm c}$) and the MTD coefficient ($b_1^{\rm t}$) as functions of the gain coefficient $\kappa$ at $\lambda=711.5$~nm (near-field resonance). The gain-assisted dielectric--Ag core-shell nanosphere parameters are the same as in Fig.~\ref{fig2}. The inset marks the gain values where ${\rm Re}(b_1^{\rm t})\approx{\rm Im}(b_1^{\rm c})\approx0$ ($\kappa\approx0.00641$) and ${\rm Re}(b_1^{\rm c})\approx0$ ($\kappa\approx0.0066$).
(b) Comparison between $(5a_2+3b_1)(3a_1+5b_2)^*$ and $b_1^{\rm t}$ at $\lambda=711.5$~nm.
 The near-linear correlation between these quantities gives rise to Eq.~(\ref{linear-correlation}).
 }
\label{fig3}
\end{figure}

With this aim, we consider a gain-assisted dielectric nanosphere of radius $r_1=125$~nm coated with a silver shell of outer radius $r_2=170$~nm in free space (Fig.~\ref{fig1}).
This core-shell configuration spatially separates plasmonic loss in the metallic shell from optical gain in the dielectric core and supports strong toroidal resonances, enabling their selective amplification, as previously observed in similar core-shell geometries~\cite{Arruda_PhysRevB109_2024}.
The core is modeled as a linear gain medium with refractive index $n_1=3.5-\imath\kappa$, where $\kappa$ is a below-threshold gain coefficient; doped AlGaAs is taken as a representative realizable gain medium operating in the linear regime~\cite{Nano,Tsakmakidis_Science339_2013}.
The Ag permittivity is described by a Drude model ${\varepsilon_{\rm Ag}(\omega)}/{\varepsilon_0}
=\varepsilon_{\rm int}-{\omega_{\rm p}^2}/[{\omega(\omega+\imath\gamma)}]$, with $\varepsilon_{\rm int}=3.7$, $\omega_{\rm p}=9.2$~eV, and $\gamma=0.02$~eV, reproducing experimental data below the interband transition threshold~\cite{Boltasseva_LaserPhotRev4_2010,Christy_PhysRevB6_1972}.
The nanosphere is illuminated by a plane wave $(\mathbf{E}_{\rm in},\mathbf{H}_{\rm in})=(E_0\hat{\mathbf{x}},H_0\hat{\mathbf{y}})e^{\imath(kz-\omega t)}$ with wavelengths $700$--$725$~nm ($1.47<kr_2<1.52$), where electric and magnetic dipole and quadrupole modes dominate the Lorenz-Mie response.

Light scattering by a core-shell nanoparticle is described by the Aden--Kerker solution of Lorenz-Mie theory~\cite{Bohren_Book_1983}.
The far-field response is characterized by the scattering efficiency $Q_{\rm sca}$,
\begin{align}
Q_{\rm sca} = \frac{2}{\rho^2}\sum_{\ell=1}^{\infty}(2\ell+1)
\left(|a_{\ell}|^2+|b_{\ell}|^2\right),
\end{align}
where $\rho=kr_2$ is the size parameter of the outer sphere and $a_{\ell}$ and $b_{\ell}$ are the electric and magnetic Lorenz-Mie coefficients of order $\ell$~\cite{Bohren_Book_1983}.

Recently, it was shown that electric and magnetic toroidal dipole excitations can be explicitly separated from spherical dipole moments within Lorenz-Mie theory~\cite{Arruda_PhysRevB109_2024}.
For moderate size parameters, the magnetic coefficient can be decomposed as
\begin{align}
b_1 \approx b_1^{\rm c} + b_1^{\rm t},
\end{align}
where $b_1^{\rm c}$ and $b_1^{\rm t}$ correspond to the Cartesian magnetic dipole (MD) and magnetic toroidal dipole (MTD), respectively.
Explicit expressions and the validity of this separation beyond the small-size limit are given in Ref.~\cite{Arruda_PhysRevB109_2024}.
The spherical magnetic dipole satisfies $\mathbf{m}(b_1)\approx \mathbf{M}(b_1^{\rm c})+k\mathbf{T}_{\rm m}(b_1^{\rm t})$, where $\mathbf{M}=\int {\rm d}^3r\,[\mathbf{r}\times\mathbf{J}(\mathbf{r})]/2$
and $\mathbf{T}_{\rm m}=-k\int {\rm d}^3r\, r^2[\mathbf{r}\times \mathbf{J}(\mathbf{r})]/20$ are the Cartesian MD and MTD, respectively.
In this work, the MTD is identified via the exact analytic separation of $b_1$, without fitting parameters [Eqs.~(44) and (45) of Ref.~\cite{Arruda_PhysRevB109_2024}].

Figure~\ref{fig2} shows the scattering efficiency and near-field intensity of a gain-assisted dielectric--Ag core-shell nanosphere, calculated via Lorenz-Mie theory~\cite{Arruda_JOpt14_2012}, as functions of wavelength and gain coefficient $\kappa$.
In Fig.~\ref{fig2}(a), far-field scattering is accurately described by the first two multipolar orders ($\ell=1,2$).
In this range, the spherical electric dipole ($a_1$), electric quadrupole ($a_2$), and magnetic quadrupole ($b_2$) remain nearly constant, while the scattering efficiency $Q_{\rm sca}$ exhibits a Fano-type lineshape dominated by the spherical magnetic dipole ($b_1$).
Decomposing $b_1$ into Cartesian MD ($b_1^{\rm c}$) and MTD ($b_1^{\rm t}$), we find partial suppression of $b_1^{\rm c}$ at $\lambda\approx711.2$~nm, coinciding with a peak in $b_1^{\rm t}$.
However, higher-order multipoles prevent the Fano dip from producing toroidal-dipole-induced invisibility ($Q_{\rm sca}\approx0$).

The magnetic Lorenz-Mie coefficient $b_1$ can be recast in a form that explicitly reveals its underlying Fano interference.
Generalizing Ref.~\cite{Tribelsky_PhysRevA93_2016} to a core-shell geometry, we obtain the exact expression
\begin{align}
b_1=\frac{F_1}{F_1+\imath G_1}=\frac{\zeta + q}{\zeta + q -\imath(\zeta q-1)},\label{b1-fano}
\end{align}
with $\zeta=[{F_1\psi_1(\rho)-G_1\chi_1(\rho)}]/[{\psi_1(n_2\rho)-B_1\chi_1(n_2\rho)}]$, $q={\chi_1(\rho)}/{\psi_1(\rho)}$,
$F_1=\psi_1'(\rho)[\psi_1(n_2\rho)-B_1\chi_1(n_2\rho)]- n_2\psi_1(\rho)[\psi_1'(n_2\rho)-B_1\chi_1'(n_2\rho)]$,
$G_1=-\chi_1'(\rho)[\psi_1(n_2\rho)-B_1\chi_1(n_2\rho)] + n_2\chi_1(\rho)[\psi_1'(n_2\rho)-B_1\chi_1'(n_2\rho)]$,
where $\psi_1$ and $\chi_1$ are the first-order Riccati-Bessel and Riccati-Neumann functions, $n_2=\sqrt{\varepsilon_2/\varepsilon_0}$ is the shell refractive index, and $B_1$, which depends on $kr_1$ and $n_1=\sqrt{\varepsilon_1/\varepsilon_0}$, is given in Ref.~\cite{Arruda_PhysRevB109_2024}.
The homogeneous-sphere result of Ref.~\cite{Tribelsky_PhysRevA93_2016} is recovered for $n_1=n_2$.

From Eq.~(\ref{b1-fano}) one obtains
\begin{equation}
|b_1|^2=\frac{1}{1+q^2}\left[\frac{(\epsilon+\overline{q})^2 + \delta^2}{\epsilon^2+1}\right],
\label{sigma-approx}
\end{equation}
which corresponds to a generalized Fano profile.
Here $\zeta=\zeta'+\imath\zeta''$, $\epsilon=\zeta'/(1+\zeta'')$, $\delta=\zeta''/(1+\zeta'')$, and $\overline{q}=q/(1+\zeta'')$. For a lossless sphere, $\zeta''(\omega)=0$, and Eq.~(\ref{sigma-approx}) reduces to a normalized Fano lineshape with asymmetry parameter $q$.
In the presence of losses and near a Fano resonance, one may further approximate $\epsilon\approx(\omega-\omega_{\rm res})/\Omega$, where $\Omega$ denotes the resonance linewidth.
Although $\zeta''$ has a cumbersome analytical form, it can be estimated at the spherical dipole resonance by imposing $\zeta'(\omega_{\rm res})=0$~\cite{Arruda_PhysRevA96_2017}.

Figures~\ref{fig2}(b) and \ref{fig2}(c) show that the MTD and Cartesian MD contributions can be actively tuned by optical gain in the dielectric core. As demonstrated in Ref.~\cite{Arruda_PhysRevB109_2024}, gain can selectively enhance the toroidal response while suppressing the Cartesian dipole at the scattering resonance.
As the gain coefficient $\kappa$ increases, the magnetic dipole contribution grows and reaches a maximum at $\kappa=0.006414$ [Fig.~\ref{fig2}(b)], corresponding to enhanced near-field magnetic response [Fig.~\ref{fig2}(e)].
This resonance may indicate the proximity of a scattering pole associated with gain compensation.
It is also accompanied by a predominantly reactive toroidal response that couples weakly to the far field, yielding a symmetric spectral lineshape.
The Fano profile in Figs.~\ref{fig2}(a) and \ref{fig2}(c) arises from interference between out-of-phase MD and MTD excitations. For all gain configurations, the MTD resonance in $Q_{\rm sca}$ coincides with enhanced magnetic-field intensity inside the core at $\lambda\approx711.5$~nm [Figs.~\ref{fig2}(d)--\ref{fig2}(f)], demonstrating near- and far-field coupling predominantly mediated by the MTD.

\begin{figure*}[htbp]
\hspace{-.6cm}
\includegraphics[width=.34\textwidth]{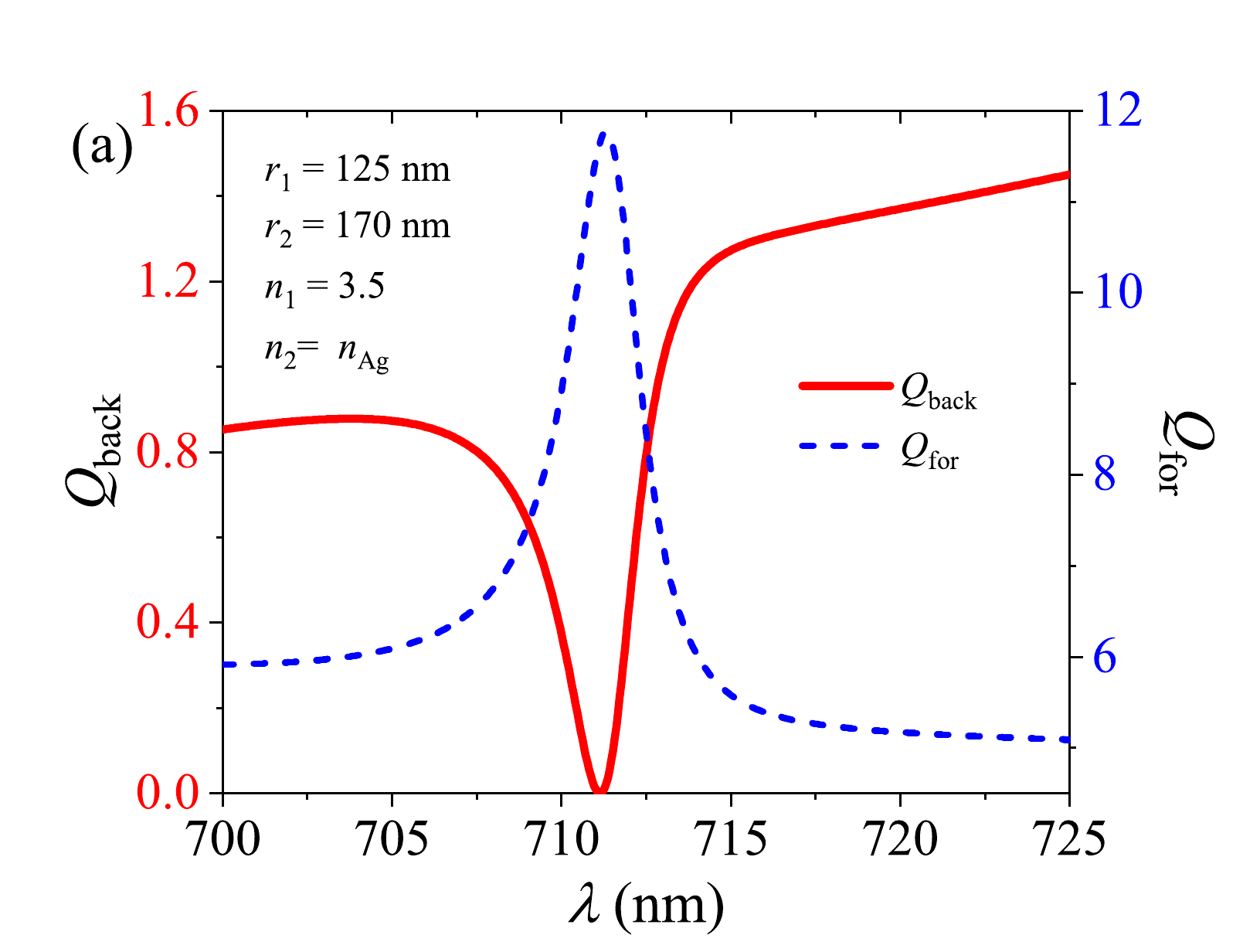}\hspace{-.6cm}
\includegraphics[width=.34\textwidth]{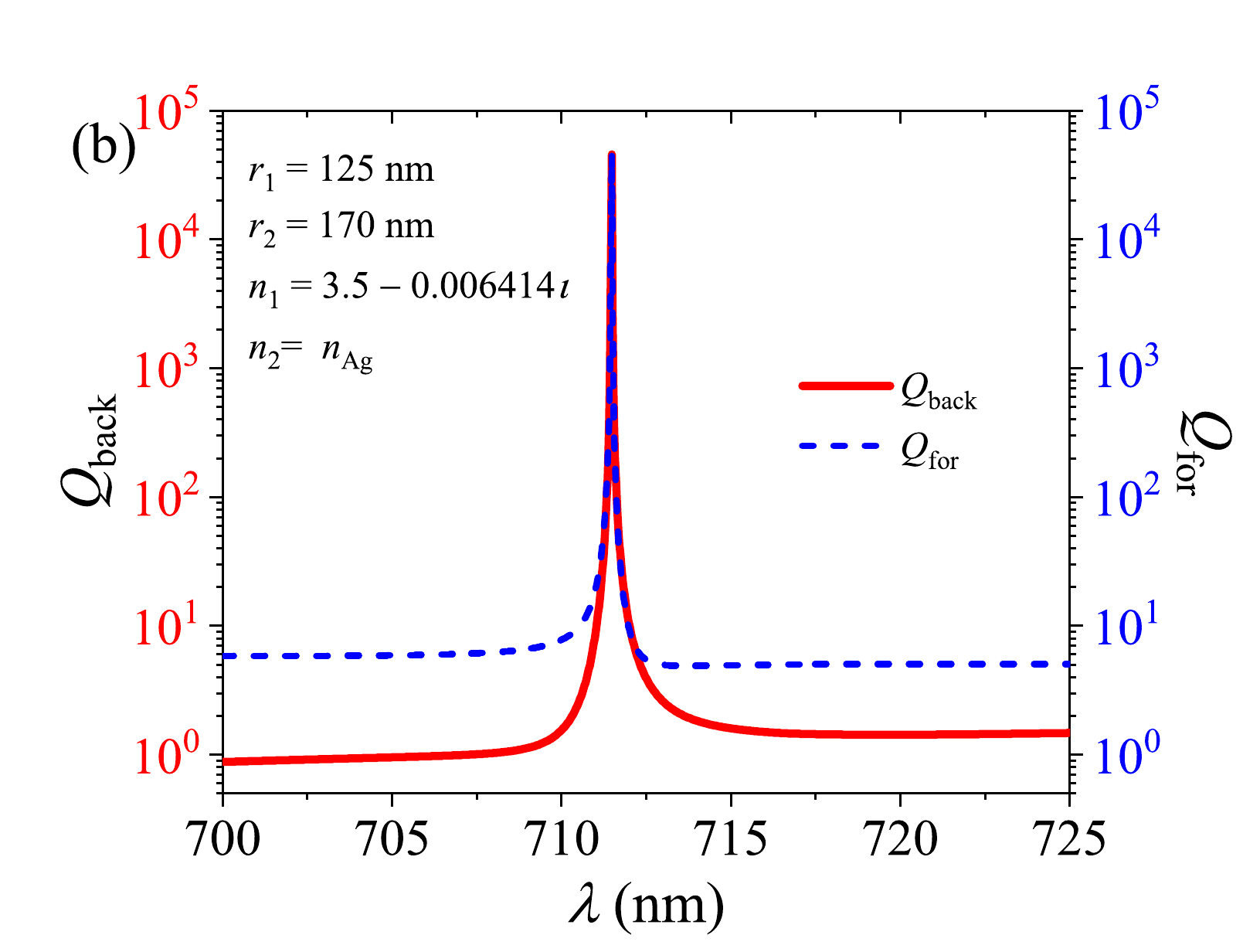}\hspace{-.6cm}
\includegraphics[width=.34\textwidth]{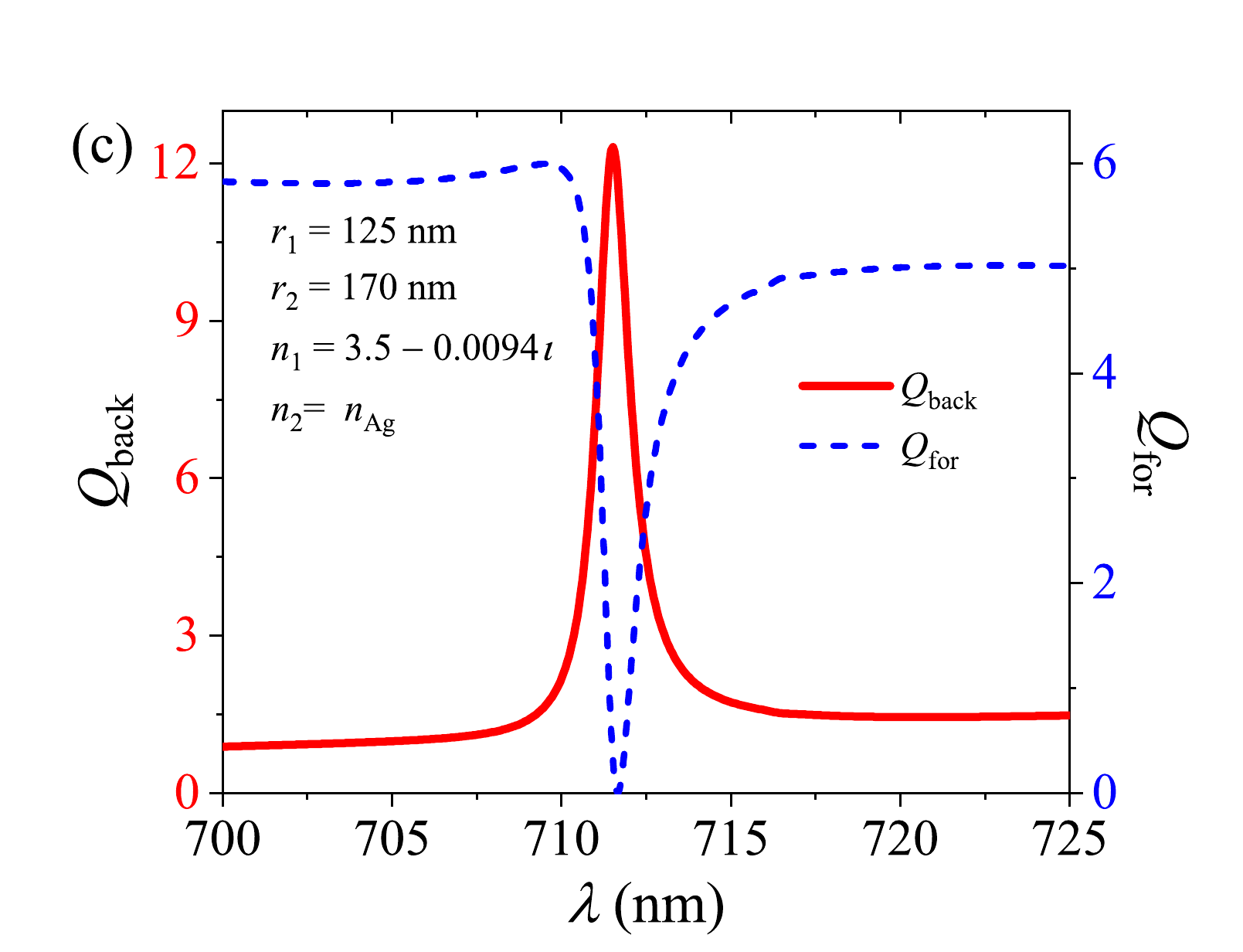}
\includegraphics[width=.36\textwidth]{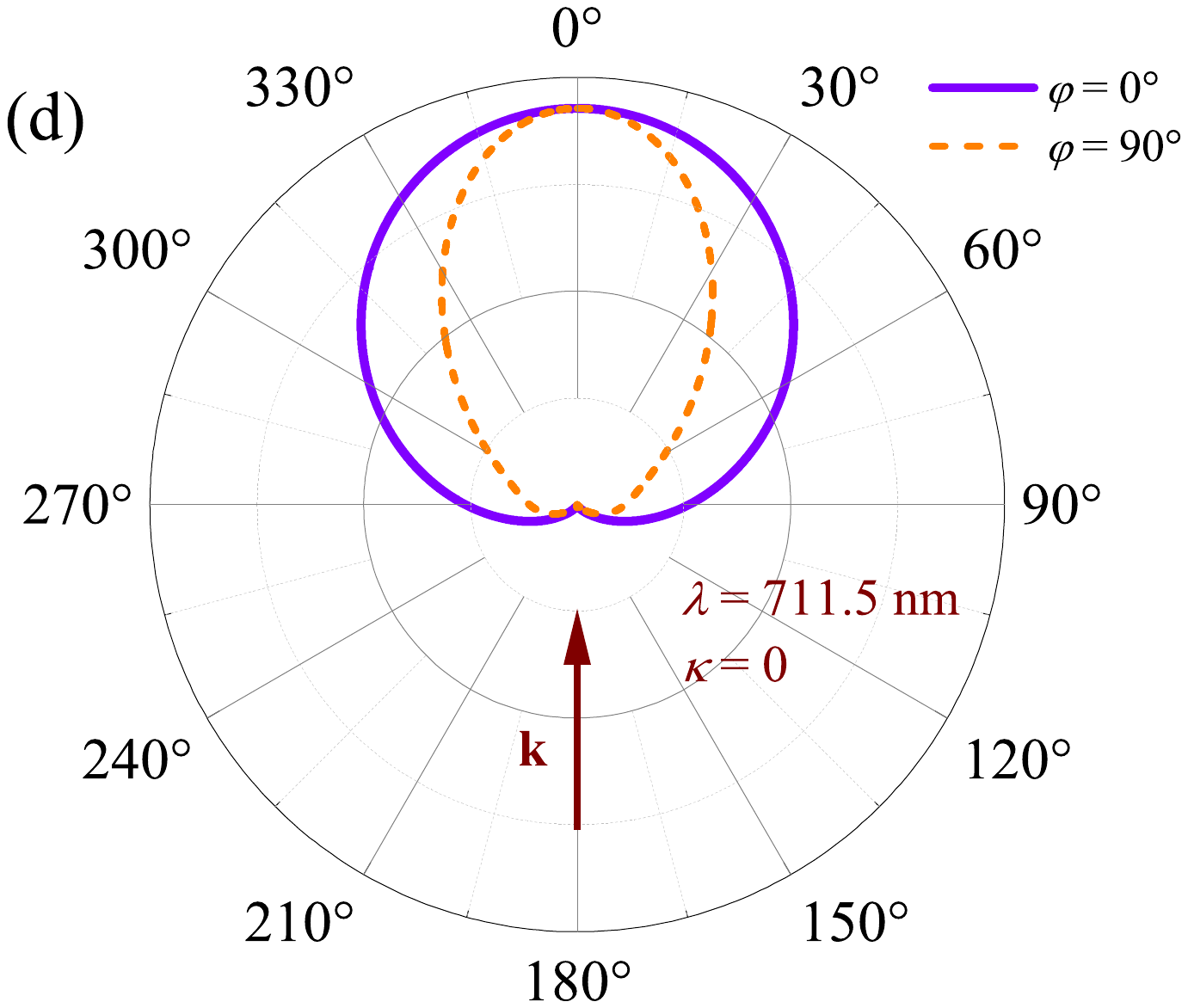}\hspace{-1.0cm}
\includegraphics[width=.36\textwidth]{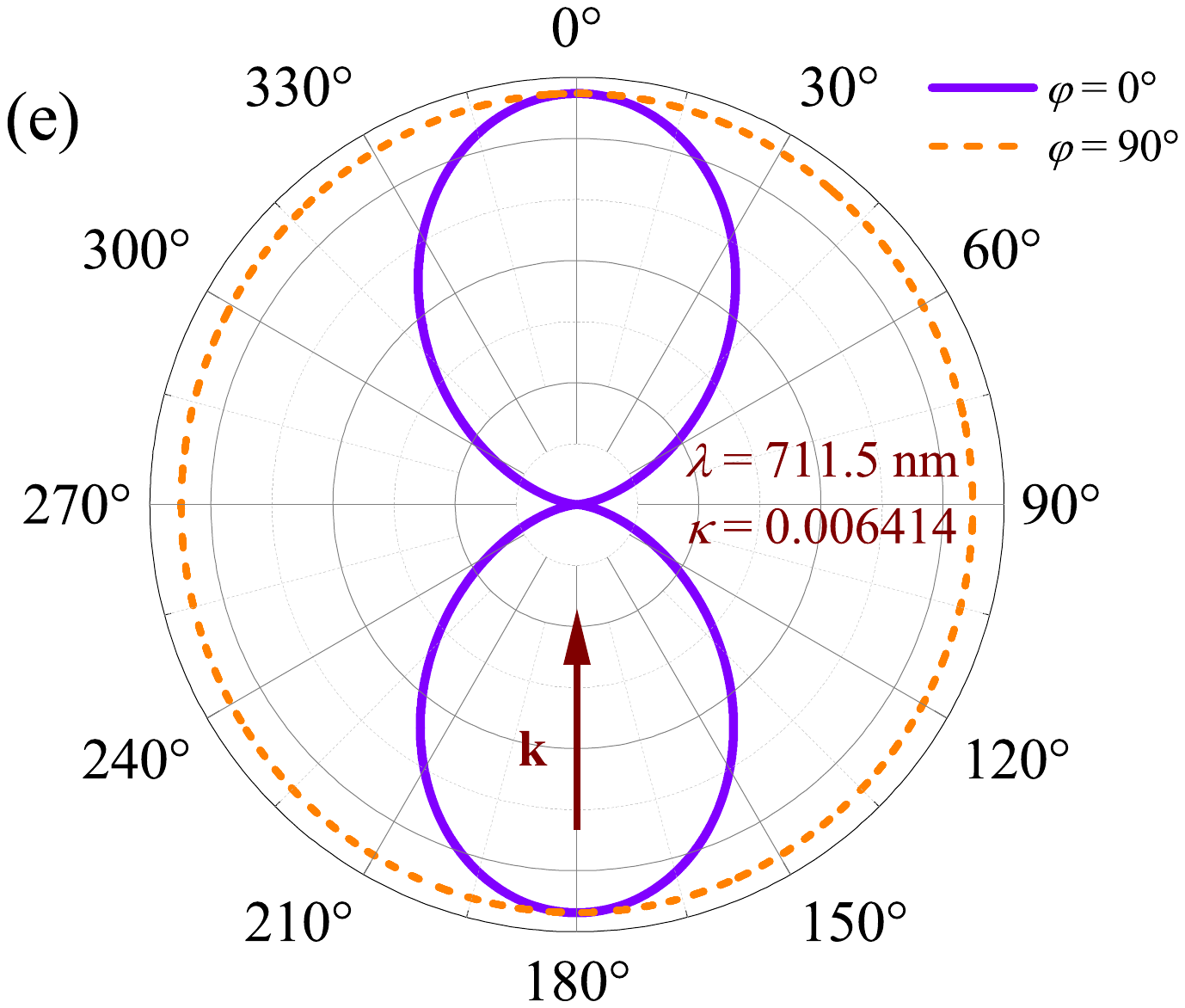}\hspace{-1.0cm}
\includegraphics[width=.36\textwidth]{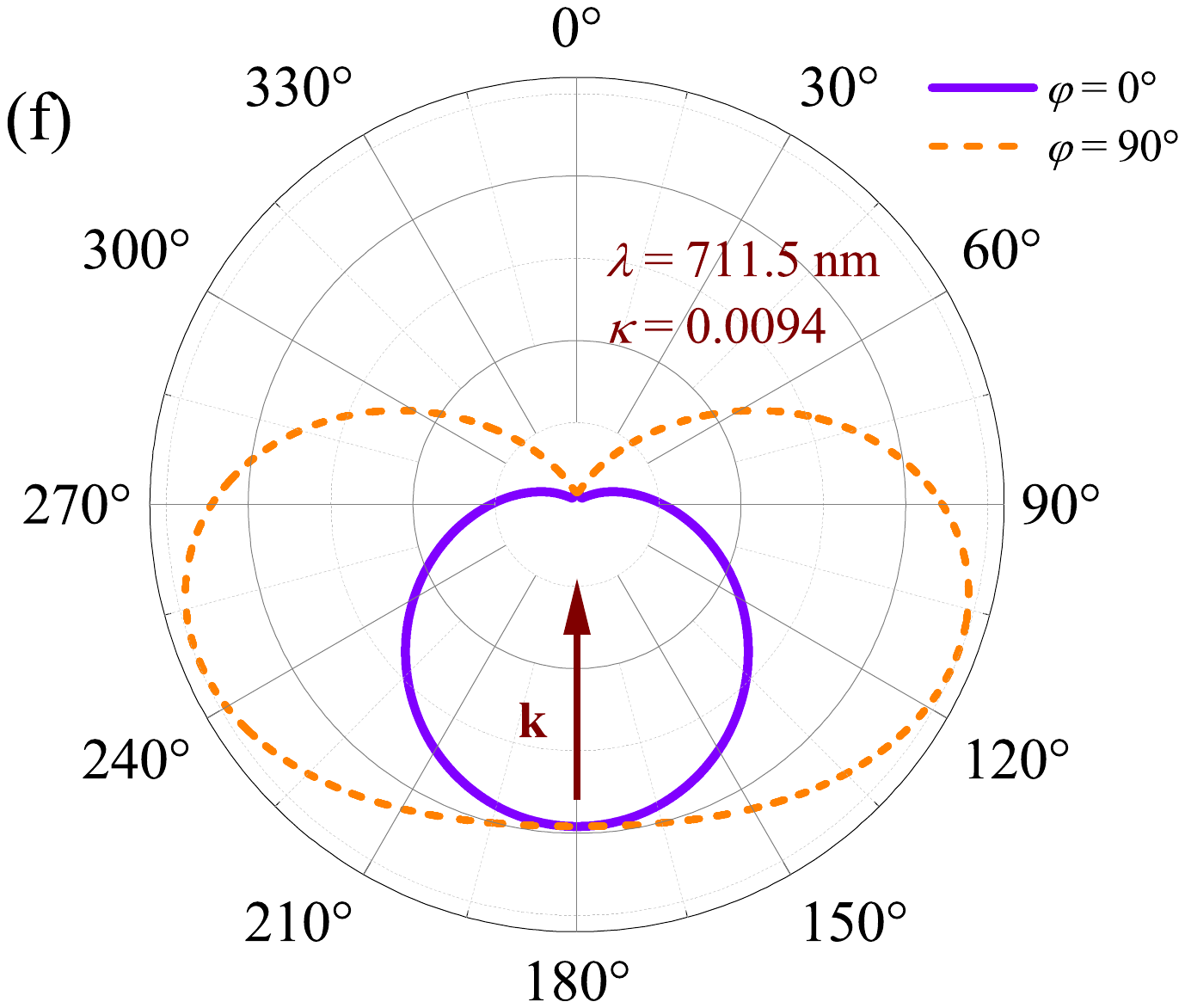}
\caption{Forward ($Q_{\rm for}$) and backward ($Q_{\rm back}$) scattering efficiencies of a gain-assisted dielectric nanosphere ($n_1=3.5-\imath\kappa$) of radius $r_1=125$~nm coated with a silver nanoshell of outer radius $r_2=170$~nm as functions of wavelength. Results are shown for (a) $\kappa=0$ (no gain), (b) $\kappa=0.006414$, and (c) $\kappa=0.0094$. Corresponding polar plots in the $xz$ ($\varphi=0^\circ$) and $yz$ ($\varphi=90^\circ$) planes at $\lambda=711.5$~nm are shown for (d) $\kappa=0$, (e) $\kappa=0.006414$, and (f) $\kappa=0.0094$.
}\label{fig4}
\end{figure*}

To examine how gain in the dielectric core modifies the MTD response in near- and far-field scattering, Fig.~\ref{fig3}(a) shows the partial coefficients $b_1^{\rm c}$ (Cartesian MD) and $b_1^{\rm t}$ (MTD) as functions of $\kappa$.
We fix $\lambda=711.5$~nm, corresponding to the wavelength of maximum field enhancement inside the core-shell nanosphere [Figs.~\ref{fig2}(d)--\ref{fig2}(f)].
The real part of $b_1^{\rm t}$ vanishes and changes sign at $\kappa\approx0.006414$, indicating a $\pi$ phase shift of the MTD relative to the Cartesian MD, while ${\rm Im}(b_1^{\rm c})$ simultaneously vanishes.
Thus, the Lorentzian resonance and near-field enhancement at $\lambda=711.5$~nm arise from ${\rm Re}(b_1^{\rm t})\approx{\rm Im}(b_1^{\rm c})\approx0$.
Since the real (imaginary) parts of Lorenz-Mie coefficients correspond predominantly to radiative (stored-energy) contributions, this configuration implies a purely reactive MTD and a purely radiative Cartesian MD.

The impact of MTD excitations is manifested in the forward ($\theta=0^{\circ}$) and backward ($\theta=180^{\circ}$) differential scattering efficiencies.
For a spherical scatterer,
\begin{align}
Q_{\pm}= \frac{1}{\rho^2}\left|\sum_{\ell=1}^{\infty}(2\ell+1)(\pm1)^{\ell}\left(a_{\ell}\pm b_{\ell}\right)\right|^2, \label{Qback}
\end{align}
where $Q_{\rm back}=Q_-$ and $Q_{\rm for}=Q_+$ are the backward and forward scattering efficiencies, respectively~\cite{Bohren_Book_1983}.
For $\rho\approx 1$, these efficiencies are well approximated by retaining $\ell\le2$,
\begin{align}
Q_{\pm} \approx \frac{1}{\rho^2}\left|(5a_2 + 3b_1)\pm (3a_1+5b_2)\right|^2.
\label{Qpm}
\end{align}

Equation~(\ref{Qpm}) shows that directional scattering arises from interference between electric and magnetic dipolar and quadrupolar modes.
In particular, excitation of a strong MTD modifies the phase and amplitude of $b_1$, enabling near-cancellation of the backward scattering amplitude under the generalized Kerker condition $5a_2 + 3b_1 \approx 3a_1 + 5b_2$, while enhancing forward scattering.
This is illustrated in Fig.~\ref{fig4}(a) for $\kappa=0$, with the polar plot in Fig.~\ref{fig4}(d) showing enhanced forward scattering at $\lambda=711.5$~nm and near-zero backward response ($Q_{\rm back}\approx0$).

Both $Q_{\rm back}$ and $Q_{\rm for}$ exhibit clear Fano lineshapes for $\kappa=0$.
In the chosen spectral range only the spherical magnetic dipole coefficient $b_1$ is resonant, whereas all other multipolar contributions remain approximately constant.
Therefore, using Eq.~(\ref{b1-fano}) and treating $a_1$, $a_2$, and $b_2$ as constants, one finds after algebraic manipulation that the forward and backward scattering amplitudes can be rewritten as
\begin{align}
&|(5a_2+3b_1)\pm(3a_1+5b_2)|^2\nonumber\\
&=\frac{1}{1+q^2}\left[
\alpha_{\pm}^2\frac{(\epsilon + q_{\alpha_{\pm}})^2}{\epsilon^2+1}
+ \beta_{\pm}^2\frac{(\epsilon + q_{\beta_{\pm}})^2}{\epsilon^2+1}
\right],
\label{Qback-fano}
\end{align}
with
$\alpha_{\pm}={\rm Re}[5a_2\pm(5b_2+3a_1)]
+ q{\rm Im}[5a_2\pm(5b_2+3a_1)] +3$,
$\beta_{\pm} = {\rm Im}[5a_2\pm(5b_2+3a_1)]
- q{\rm Re}[5a_2\pm(5b_2+3a_1)]$,
$q_{\alpha_{\pm}}=(3\overline{q}-\beta_{\pm})/\alpha_{\pm}$,
and
$q_{\beta_{\pm}}=\alpha_{\pm}/\beta_{\pm}-{3}/[({1+\zeta''})\beta_{\pm}]$.
The upper (lower) sign corresponds to the forward (backward) scattering direction, $\epsilon$ is the reduced frequency detuning from resonance, and $q_{\alpha_{\pm}}$ and $q_{\beta_{\pm}}$ are generalized Fano asymmetry parameters.

Equation~(\ref{Qback-fano}) shows that forward and backward scattering can be interpreted as an incoherent superposition of two effective Fano channels, weighted by $\alpha_{\pm}^2$ and $\beta_{\pm}^2$ and characterized by asymmetry parameters $q_{\alpha_{\pm}}$ and $q_{\beta_{\pm}}$.
These channels result from interference between a resonant spherical magnetic
dipole and a slowly varying background formed by nonresonant electric and magnetic multipoles.
Directional suppression of backscattering ($Q_{\rm back}\approx 0$) occurs when $\epsilon+q_{\alpha_-}=0$, since the second term in Eq.~(\ref{Qback-fano}) remains finite and contributes only as background.
Numerical calculations confirm that directional scattering is governed by the balance between $\alpha_{\pm}$ and $q_{\alpha_{\pm}}$, encoding interference between conventional multipoles and toroidal excitations.

For a gain-assisted dielectric core with $\kappa=0.006414$, Fig.~\ref{fig4}(b) shows that both $Q_{\rm back}$ and $Q_{\rm for}$ are enhanced at $\lambda\approx711.5$~nm.
At this wavelength, the MTD couples weakly to the far field, yielding a predominantly magnetic-dipolar pattern [Fig.~\ref{fig4}(e)].
The condition ${\rm Re}(b_1^{\rm t})\approx{\rm Im}(b_1^{\rm c})\approx0$ coincides with ${\rm Re}[(5a_2+3b_1)(3a_1 + 5b_2)^{*}] \approx 0$, leading to $Q_{\rm back}\approx Q_{\rm for}$.
In this regime, the spectra exhibit Lorentzian lineshapes, corresponding to $\zeta''\approx -1$, where the generalized Fano parameters diverge and the response becomes purely resonant.

For $\kappa=0.0094$, Fig.~\ref{fig4}(c) shows near-zero forward scattering ($Q_{\rm for}\approx0$) with enhanced backscattering at $\lambda\approx711.7$~nm.
The generalized Kerker condition reads $5a_2 + 3b_1 \approx -3a_1 - 5b_2$, corresponding to the Fano dip
$\epsilon+q_{\alpha_+}=0$.
Gain-induced amplification of the MTD enables this balance even in the presence of higher-order multipoles, yielding strongly directional backscattering [Fig.~\ref{fig4}(f)].

Remarkably, at the near-field resonance wavelength ($\lambda\approx711.5$~nm), we observe an approximate proportionality between the MTD coefficient and the interference term controlling the forward and backward scattering amplitudes,
\begin{align}
b_1^{\rm t}\approx\frac{1}{7}(5a_2+3b_1)(3a_1+5b_2)^{*},\label{linear-correlation}
\end{align}
as shown in Fig.~\ref{fig3}(b).
This relation is obtained numerically from the near-linear correlation shown in Fig.~\ref{fig3}(b) for the specific geometry and material parameters considered here and establishes a direct link between toroidal phase reversal and the multipolar interference responsible for directional scattering.
Using this relation, the generalized Kerker conditions in the present system can be written as
\begin{align}
&Q_{\pm}\approx 0:\ {\rm Re}(b_1^{\rm t}) \approx \mp\frac{1}{7}|3a_1+5b_2|^2,\quad {\rm Im}(b_1^{\rm t})\approx0;\\
&Q_{+}\approx Q_{-}:\ {\rm Re}(b_1^{\rm t})\approx 0.
\end{align}
This provides a clear physical interpretation of toroidal-dipole-induced directionality as a generalized Kerker condition mediated by higher-order multipolar interference.

Although optimal backward and forward suppression occur at slightly different wavelengths for distinct gain values ($711.2$ and $711.7$~nm, respectively), we fix the excitation wavelength at $\lambda=711.5$~nm in Figs.~\ref{fig4}(d)--\ref{fig4}(f) to isolate the gain-induced MTD phase evolution. This choice corresponds to the near-field intensity resonance and ensures that the directional transition is driven by gain-controlled interference rather than spectral detuning. The selected gain coefficients illustrate the mechanism and are representative of effective linear gain values reported for dielectric and semiconductor media at visible and near-infrared wavelengths~\cite{Nano,Tsakmakidis_Science339_2013}.

\begin{figure*}[htbp]
\includegraphics[width=.28\textwidth]{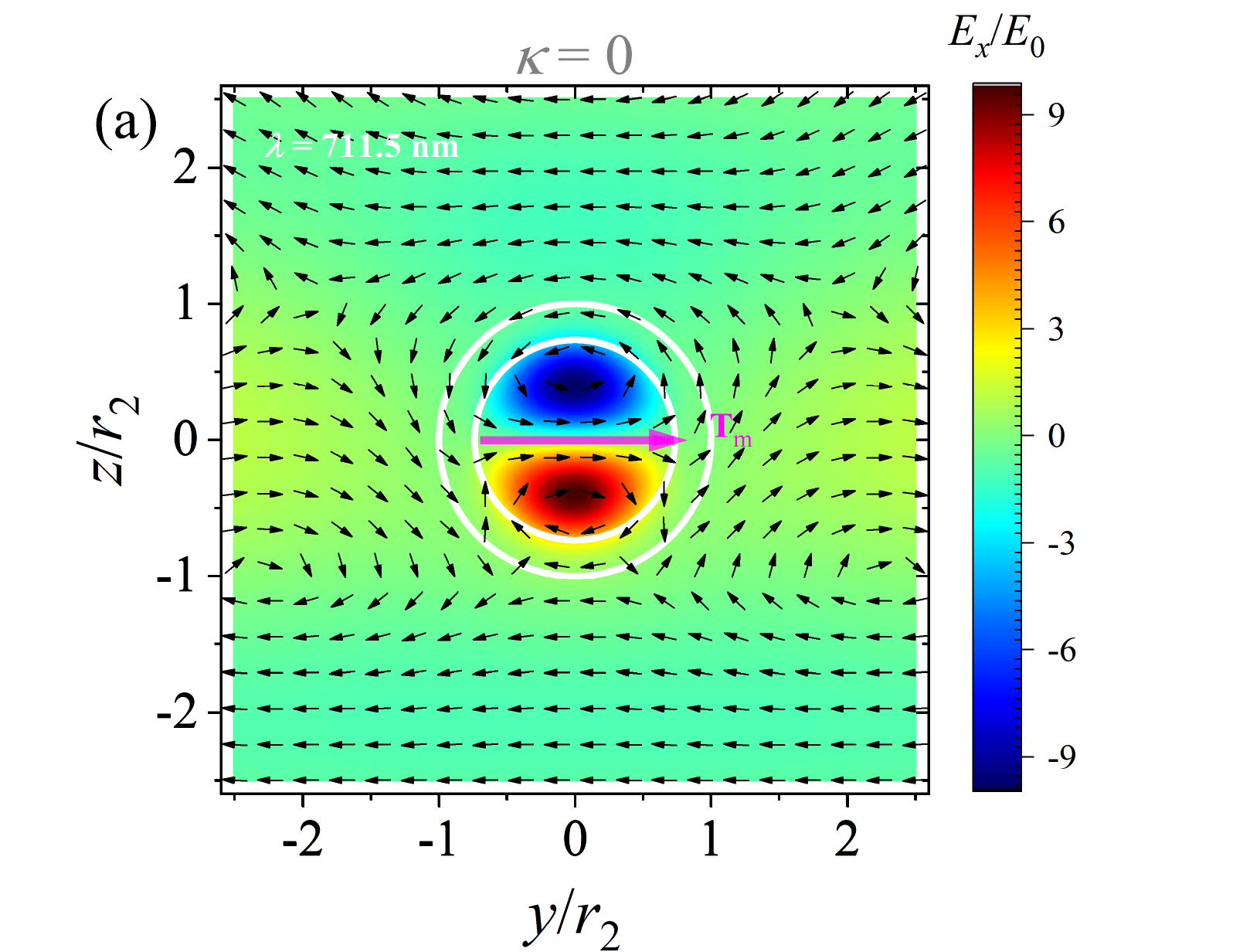}\hspace{-.9cm}
\includegraphics[width=.28\textwidth]{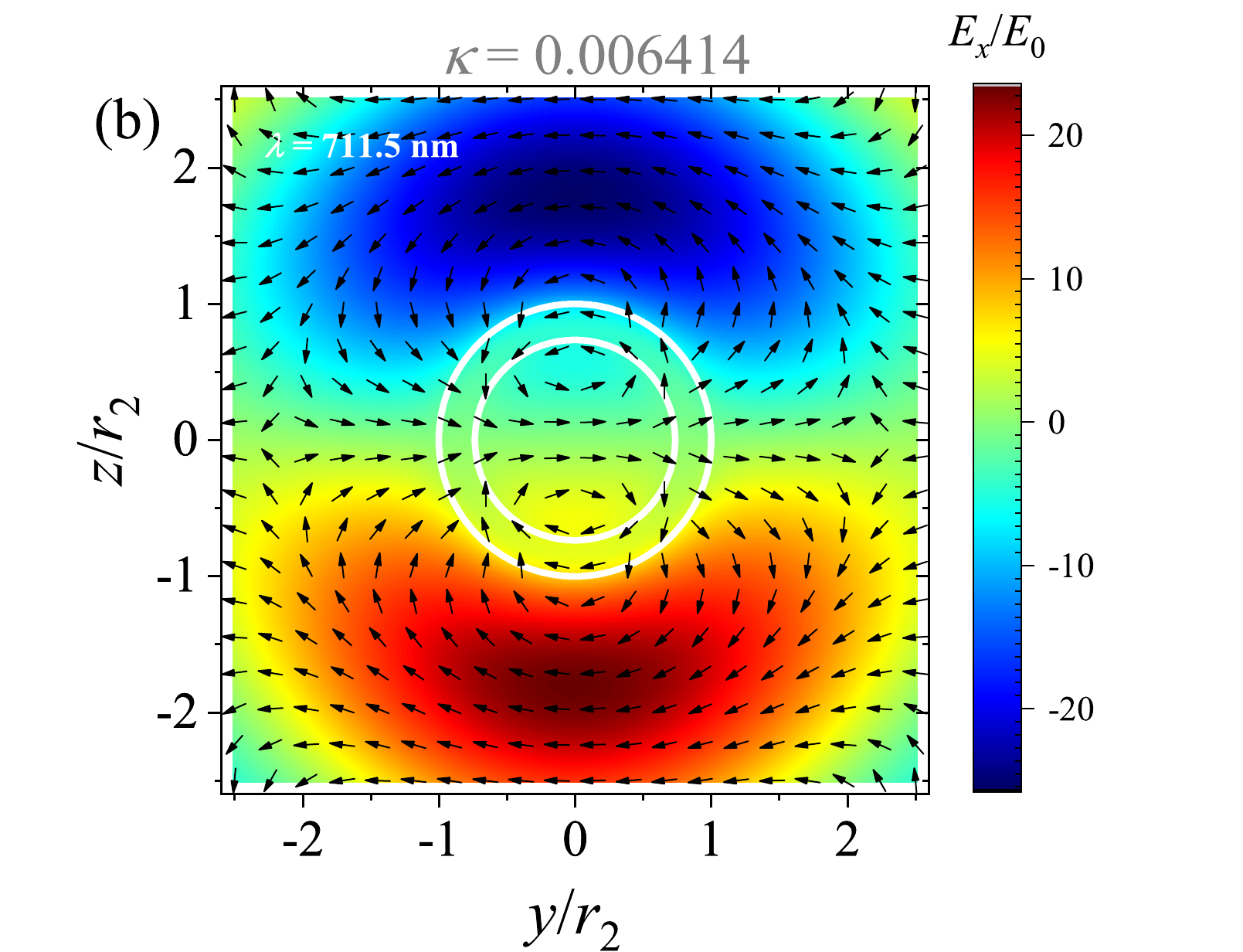}\hspace{-.9cm}
\includegraphics[width=.28\textwidth]{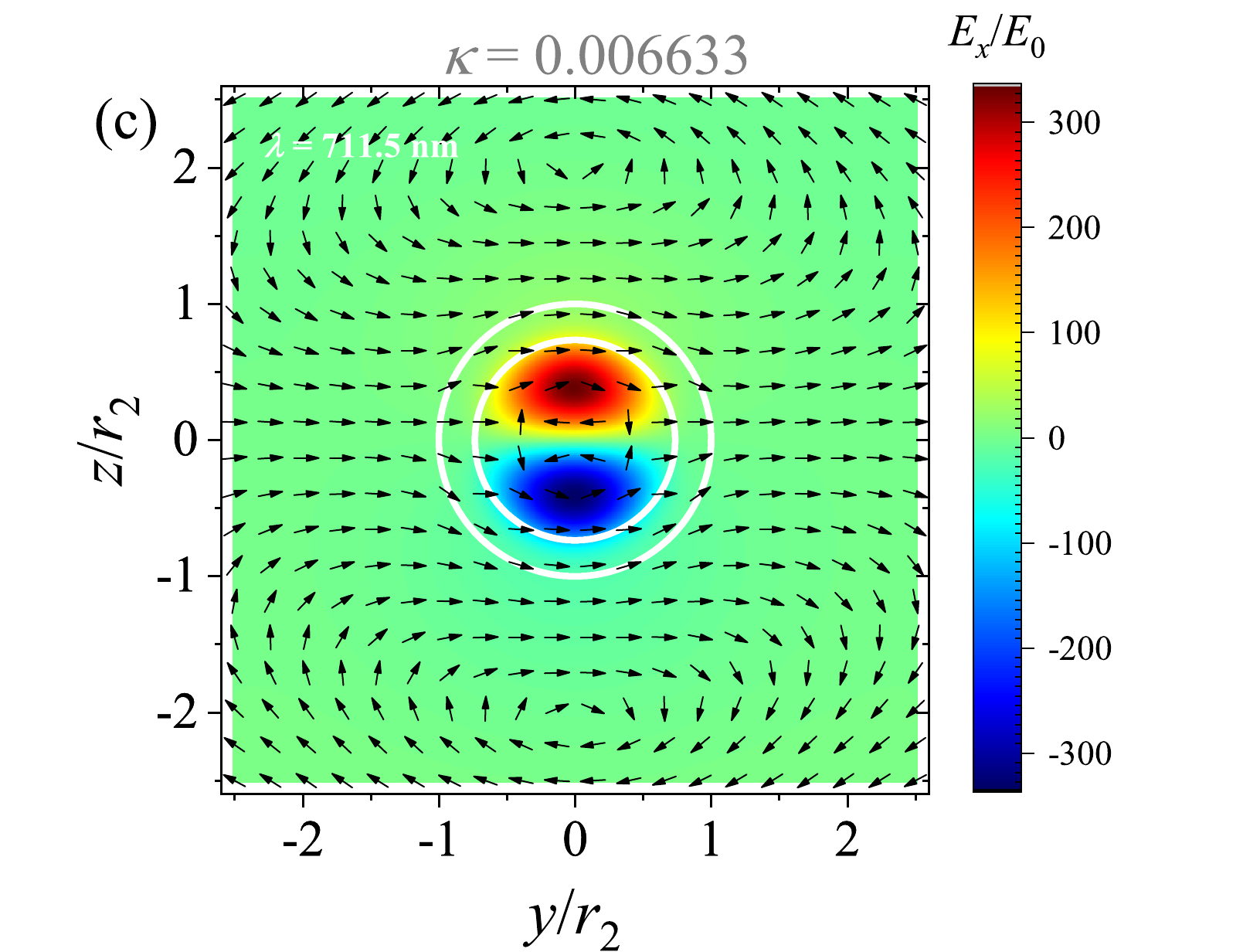}\hspace{-.9cm}
\includegraphics[width=.28\textwidth]{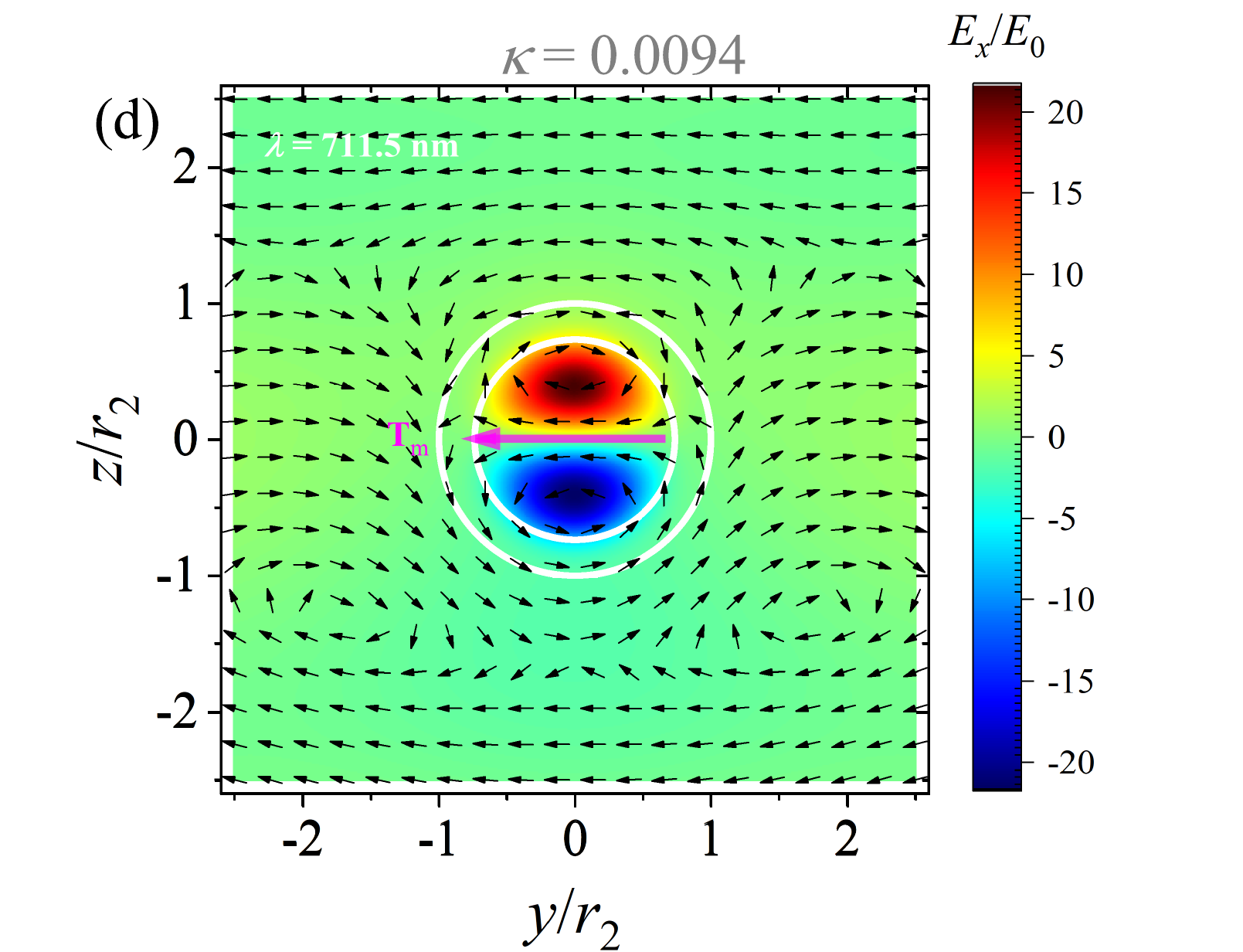}\hspace{-.9cm}
\includegraphics[width=.28\textwidth]{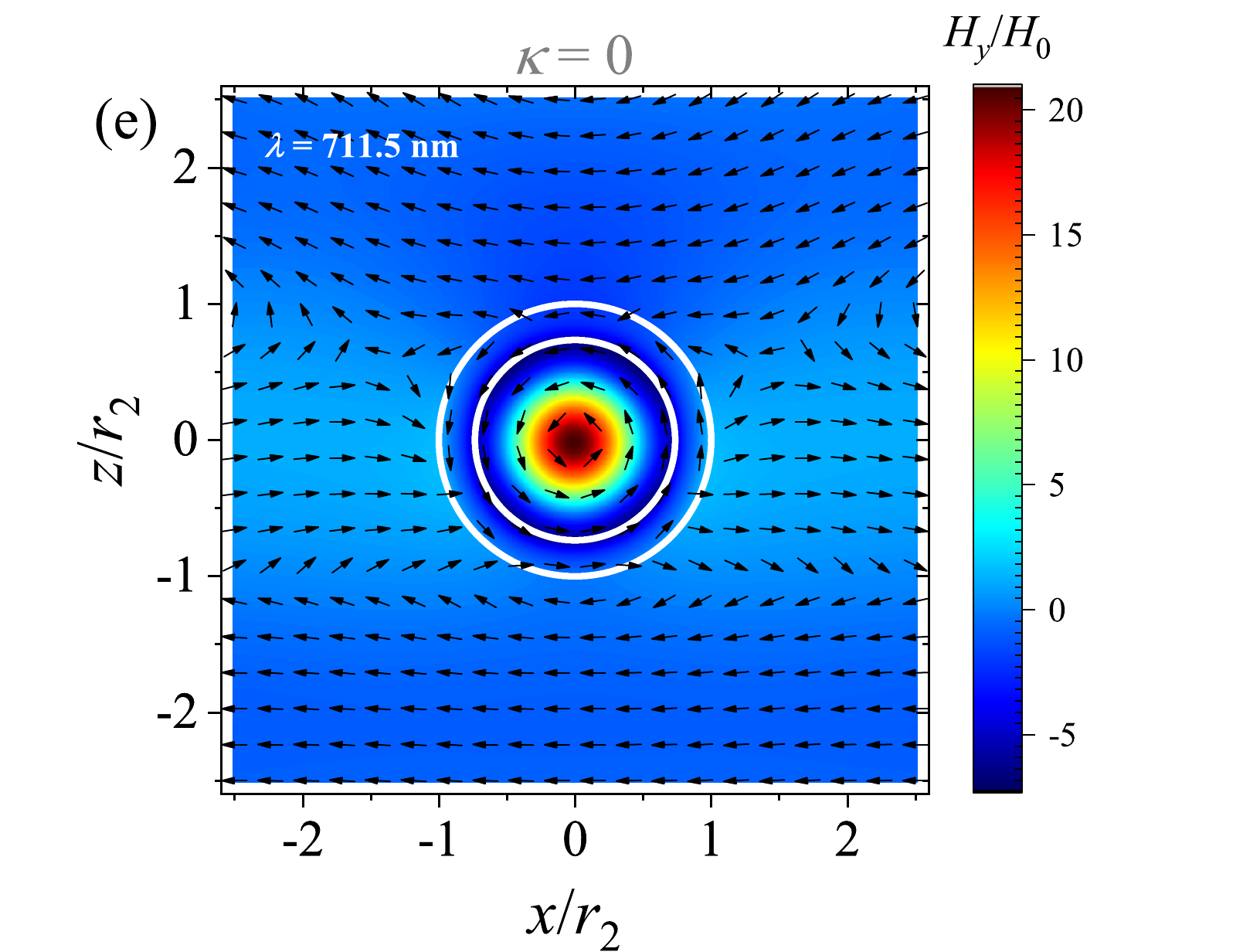}\hspace{-.9cm}
\includegraphics[width=.28\textwidth]{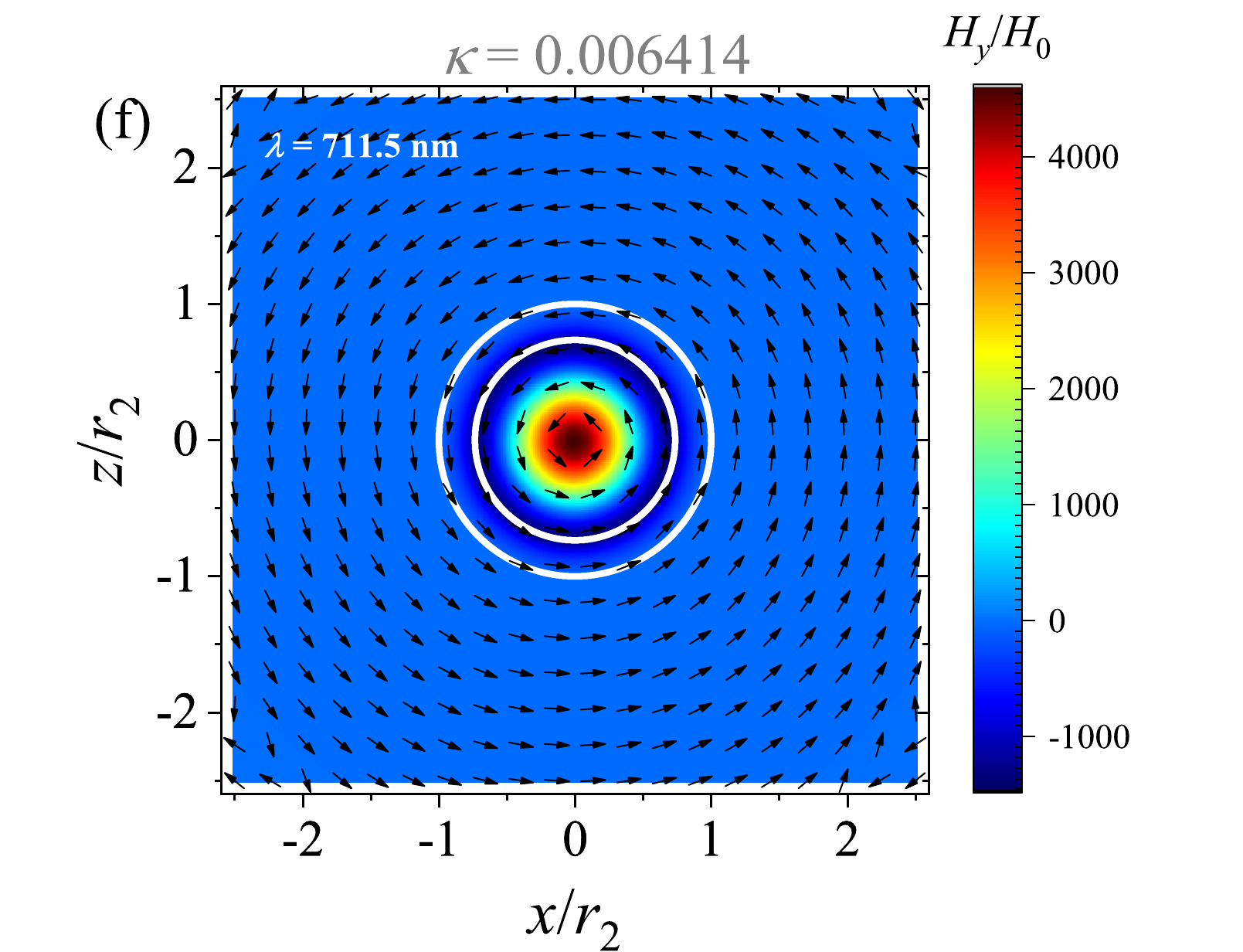}\hspace{-.9cm}
\includegraphics[width=.28\textwidth]{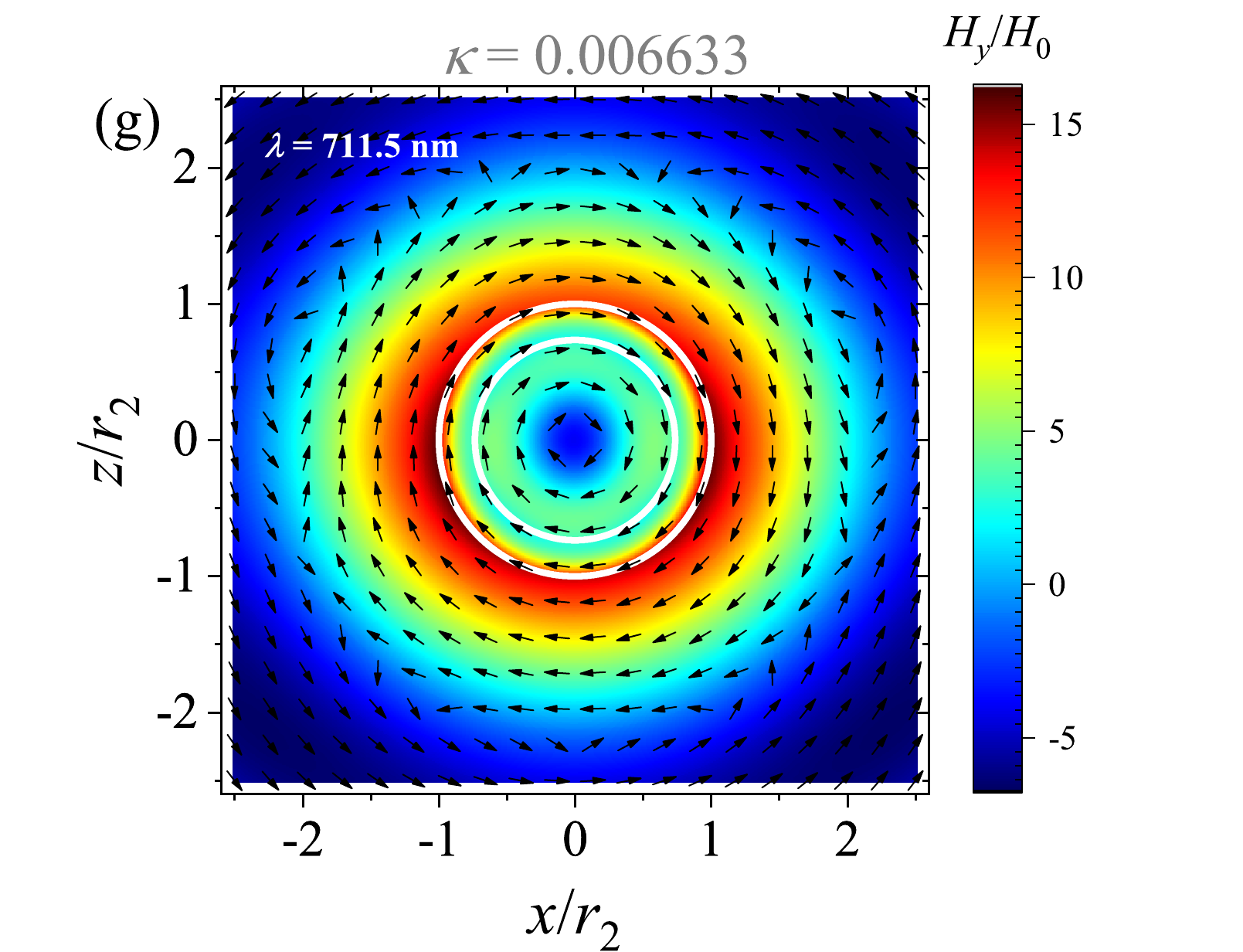}\hspace{-.9cm}
\includegraphics[width=.28\textwidth]{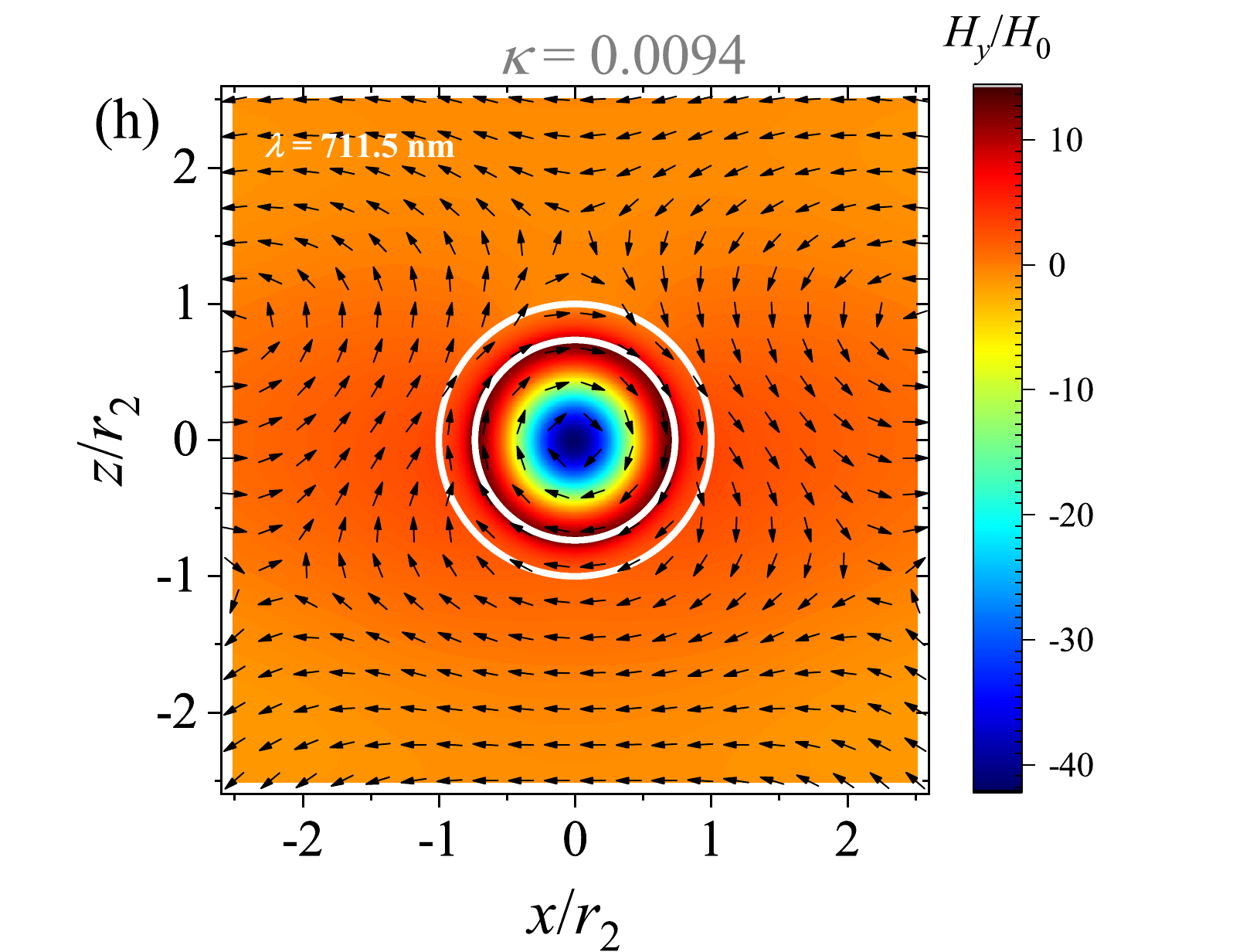}
\caption{
Electric and magnetic near-field distributions in planes crossing the center of a gain-assisted nanosphere ($n_1=3.5-\imath\kappa$) of radius $r_1=125$~nm coated with a silver nanoshell of outer radius $r_2=170$~nm at the near-field intensity resonance, $\lambda=711.5$~nm. Color maps of $E_x$ and vector plots of normalized $\mathbf{H}$ in the $yz$ plane are shown for (a) $\kappa=0$ (no gain), (b) $\kappa=0.006414$, (c) $\kappa=0.006633$, and (d) $\kappa=0.0094$. Color maps of $H_y$ and vector plots of normalized $\mathbf{E}$ in the $xz$ plane are shown in (e)--(h) for the same gain values. Color scales differ between panels because of variations in field intensity.
}\label{fig5}
\end{figure*}

To further analyze the effect of gain, Fig.~\ref{fig5} shows the near-field distributions of $E_x$ and $H_y$ in planes parallel to the incident propagation direction ($z$ axis) and crossing the center of the core-shell nanosphere. For $\kappa=0$, where only the MTD is resonant at $\lambda\approx711$~nm [Fig.~\ref{fig2}(a)], Fig.~\ref{fig5}(a) ($yz$ plane) exhibits the characteristic near-field pattern of a pure MTD excitation, with circulating magnetic-field lines around regions of enhanced electric field, forming a toroidal pattern in Fig.~\ref{fig5}(e) ($xz$ plane).
For $\kappa=0.006414$, corresponding to ${\rm Re}(b_1^{\rm t})\approx{\rm Im}(b_1^{\rm c})\approx0$, the MTD remains strongly excited, although the resulting radiation pattern resembles that of a conventional magnetic dipole [Figs.~\ref{fig5}(b) and \ref{fig5}(f)] because of interference with the Cartesian MD contribution.

In Figs.~\ref{fig5}(c) and \ref{fig5}(g), for $\kappa=0.006633$, the $E_x$ and $H_y$ fields inside the sphere change sign, indicating a reorganization of the toroidal current distribution in the near field.
This occurs when ${\rm Re}(b_1^{\rm c})\approx0$ [inset of Fig.~\ref{fig3}(a)].
As the gain further increases, the MTD undergoes a progressive rearrangement in which the
circulation direction of the underlying poloidal currents reverses.
For $\kappa = 0.0094$, Figs.~\ref{fig5}(d) and \ref{fig5}(h) show a MTD excitation with
opposite handedness, corresponding to a reversal of the magnetic-field circulation and of the underlying poloidal currents.
This change results from the gain-induced phase evolution of $b_1^{\rm t}$ and $b_1^{\rm c}$.

In conclusion, we have shown that magnetic toroidal dipoles can simultaneously enable near-field confinement and far-field directionality in gain-assisted dielectric--metal core-shell nanospheres. Optical gain selectively enhances the toroidal mode by compensating intrinsic losses, enabling interference regimes inaccessible in passive structures. In the absence of gain, destructive interference between Cartesian MD and MTD excitations suppresses backscattering and produces Fano lineshapes in the scattering spectra.
As the gain coefficient $\kappa$ increases, gain-induced phase evolution drives ${\rm Im}(b_1^{\rm c})$ and ${\rm Re}(b_1^{\rm t})$ through zero around $\kappa_{\rm c}$, yielding a scattering response dominated by magnetic-dipole radiation with symmetric forward and backward scattering.
For $\kappa>\kappa_{\rm c}$, the toroidal phase reverses, shifting destructive interference to the forward direction and producing dominant backscattering. This transition corresponds to a sign reversal of the magnetic toroidal moment and reconfiguration of the underlying poloidal currents, demonstrating active switching of directional scattering.
Our findings establish magnetic toroidal dipoles as an effective control knob for directional scattering, with implications for active nanoantennas and non-Hermitian nanophotonics.

%

\end{document}